\documentclass[10pt,twocolumn,superscriptaddress,nofootinbib]{revtex4-1}

\usepackage{amsmath}
\usepackage{amssymb}
\usepackage{graphicx}
\usepackage{color}
\usepackage{tikz}
\usepackage{url}
\usepackage{xr}

\begin{document}

\title{Compensating for population  sampling in simulations of epidemic spread on temporal contact networks}
\date{\today}
\author{Mathieu G\'enois}
\affiliation{Aix Marseille Universit\'e, Universit\'e de Toulon, CNRS, CPT, UMR 7332, 13288 Marseille, France}
\author{Christian L. Vestergaard}
\affiliation{Aix Marseille Universit\'e, Universit\'e de Toulon, CNRS, CPT, UMR 7332, 13288 Marseille, France}
\author{Ciro Cattuto}
\affiliation{Data Science Laboratory, ISI Foundation, Torino, Italy}
\author{Alain Barrat}
\affiliation{Aix Marseille Universit\'e, Universit\'e de Toulon, CNRS, CPT, UMR 7332, 13288 Marseille, France}
\affiliation{Data Science Laboratory, ISI Foundation, Torino, Italy}

\begin{abstract} 
Data describing human interactions often suffer from incomplete sampling of the underlying population.
As a consequence, the study of contagion processes using data-driven models can lead to a severe underestimation of the epidemic risk.
Here we present a systematic method to alleviate this issue and obtain a better estimation of the risk in the context of epidemic models informed by high-resolution time-resolved contact data.
We consider several such data sets collected in various contexts and perform controlled resampling experiments.
We show how the statistical information contained in the resampled data can be used to build a series of surrogate versions of the unknown contacts.
We simulate epidemic processes on the resulting reconstructed data sets and show that it is possible to obtain good estimates of the outcome of simulations performed using the complete data set.
We discuss limitations and potential improvements of our method.
\end{abstract}

\maketitle

Human interactions play an important role in determining the potential transmission routes of infectious diseases and other contagion phenomena~\cite{BBV}.
Their measure and characterisation thus represent an invaluable contribution to the study of transmissible diseases  \cite{Read:2012}.
While surveys and diaries in which volunteer participants record  their encounters
\cite{Edmunds:1997,Read:2008,Mossong:2008,Danon:2012,Danon:2013} have provided crucial insights (see however
\cite{Read:2008,Smieszek:2012,Smieszek:2014} for recent investigations of the  memory biases
inherent in self-reporting procedures), new approaches have recently emerged
to measure contact patterns between individuals with high resolution,
using wearable sensors that can detect the proximity of other similar devices
\cite{Hui:2005,ONeill:2006,Eagle:2009,Vu:2010,Salathe:2010,Hashemian:2010,Cattuto:2010,Hornbeck:2012,Stopczynski:2014,Obadia:2015,Toth:2015}.
The resulting measuring infrastructures register contacts specifically within the closed
population formed by the participants wearing sensors, with typically high spatial and temporal resolutions. In the recent years, several
data gathering efforts have used such methods to obtain, analyse and
publish data sets describing the contact patterns between individuals in various contexts
in the form of temporal networks \cite{Salathe:2010,Isella:2011,Stehle:2011,Stehle:2011a,Fournet:2014,Toth:2015}:
 nodes represent individuals and, at each time step,
a link is drawn between pairs of individuals who are in contact \cite{Holme:2012}.
Such data has been used to inform models of epidemic spreading phenomena used
to evaluate epidemic risks and mitigation strategies
in specific, size-limited contexts such as schools or hospitals
\cite{Salathe:2010,Stehle:2011,Lee:2012,Machens:2013,Smieszek:2013,Chowell:2013,Masuda:2013,Gemmetto:2014,Voirin:2015,Obadia:2015,Toth:2015},
finding in particular outcomes consistent with observed outbreak data \cite{Toth:2015} or providing evidence of links between specific contacts
and transmission events \cite{Voirin:2015,Obadia:2015}.

Despite the relevance and interest of such detailed data sets, as illustrated by
these recent investigations, they suffer from the intrinsic limitation
of the data gathering method: contacts are registered only between participants wearing sensors.
Contacts with and between individuals who do not wear sensors are thus missed.
In other words, as most often not all individuals accept to participate by wearing sensors,
many data sets obtained by  such techniques suffer from population sampling,
despite efforts to maximise participation through e.g. scientific engagement of participants \cite{Conlan:2011,Fournet:2014}.
Hence, the collected data only contains information on contacts occurring among a fraction of the population under study.

Population sampling is well-known to affect the properties of static networks \cite{Granovetter:1976,Frank:1978,Achlioptas:2005}:
various statistical properties and mixing patterns of the contact network of a fraction of the population of interest may differ
from those of the whole population, even
if the sampling is uniform \cite{Kossinets:2006,Ghani:1998a,Ghani:1998b,Onnela:2012},
and several works have focused on inferring
network statistics from the knowledge of incomplete network data \cite{Ghani:1998b,Viger:2007,Bliss:2014,Kolaczyk:arx,Cimini:arx}.
Both structural and temporal properties of time-varying networks might as well be affected by missing data effects \cite{Ghani:1998b,Cattuto:2010}.

In addition, a crucial though little studied consequence of such missing data is that simulations of
dynamical processes in data-driven models can be affected if incomplete data are used \cite{Ghani:1998a,Ghani:1998b,Bobashev:2008}.
For instance, in simulations of epidemic spreading, excluded nodes
are by definition unreachable and  thus equivalent to immunised nodes.
Due to herd vaccination effects, the outcome of  simulations of epidemic models
on sampled networks is thus expected to be underestimated with respect to simulations on the whole
network. (We note however, that in the different context of transportation networks,
it was found in  \cite{Bobashev:2008} that the inclusion of the most important transportation nodes can be sufficient to
describe the global worldwide spread of influenza-like illnesses, at least in terms of times of arrival of the spread in various cities.)
How to estimate the outcome of dynamical processes on contact networks using incomplete data remains an open question.

Here we make progresses on this issue for incompletely sampled data describing networks of human face-to-face interactions,
collected by infrastructures based on sensors,  under the assumption
that the population participating to the data collection is a uniform random sample of the whole
population of interest. (We do not therefore address here the issue of non-uniform sampling of contacts that may result
from other measurement methods such as diaries or surveys.)
We proceed through resampling experiments on empirical data sets in which we exclude
uniformly at random a fraction of the individuals (nodes of the contact network). We
measure how relevant network statistics vary under such uniform resampling and confirm
that, although some crucial properties are stable, numerical simulations of spreading processes performed
using incomplete data lead to strong underestimations of the epidemic risk. Our goal and main contribution
consists then in putting forward and comparing a hierarchy of systematic methods to provide better estimates of the outcome of
models of epidemic spread in the whole population under study.
To this aim, we do not try to infer the true sequence of missing contacts. Instead, the
methods we present consist in the construction of surrogate contact sequences for
the excluded nodes, using only structural and temporal information available in the resampled contact data.
We perform simulations of spreading processes on the reconstructed data sets, obtained by the union of the
resampled and surrogate contacts,  and investigate how their outcomes vary depending on the amount
of information included in the reconstruction method. We show that it is possible
to obtain outcomes close to the results  obtained on the complete
data set, while, as mentioned above, using only the incomplete data severely underestimates the epidemic risk.
We show the efficiency of our procedure using three data sets
collected in widely different contexts and representative of very different population structures found in day-to-day life:
a scientific conference, a high school and a workplace.
We finally discuss the limitations of our method in terms of sampling range, model parameters and population sizes.

\section*{Results}

\subsection*{Data and Methodology}

We consider data sets describing contacts between individuals, collected by the  SocioPatterns collaboration (\url{http://www.sociopatterns.org})
in three different settings: a workplace (office building, InVS) \cite{Genois:2015}, a high school (Thiers13) \cite{Fournet:2014} and
a scientific conference (SFHH) \cite{Isella:2011,Stehle:2011}. These data correspond to the
close face-to-face proximity of individuals equipped with wearable sensors, at a temporal resolution of $20$ seconds~\cite{Cattuto:2010}.
Table~\ref{tab:data} summarises the characteristics of each data set.
The contact data are represented by temporal networks, in which nodes represent the participating individuals and a link
between two nodes $i$ and $j$ at time $t$ indicates that the two corresponding persons were in contact at that time.
These three data sets were chosen as representative of different types of day-to-day contexts and of different contact
network structures: the SFHH data correspond to a rather homogeneous contact network;
the InVS and Thiers13 populations were instead structured in departments and classes, respectively. Moreover,
high school classes (Thiers13) are of similar sizes while the InVS department sizes are unequal. Finally,
the high school contact patterns (Thiers13) are constrained by strict and repetitive school schedules, while
contacts in offices are less regular across days.

To quantify how the incompleteness of data, assumed to stem from
a uniformly random participation of individuals to the data collection,
affects  the outcome of simulations of dynamical processes,
we consider as ground truth the available data and perform population resampling experiments by removing a fraction $f$ of the nodes uniformly at random. (Note that the full data sets are also samples of all the contacts that occurred in the populations, as the participation rate was lower than $100\%$ in each case. In the Thiers13 case however, the participation
rate was quite high.)
We then simulate on the resampled data
the paradigmatic Susceptible-Infectious-Recovered (SIR) and the
Susceptible-Infectious-Susceptible (SIS) models of epidemic propagation.
In these models, a susceptible (S) node becomes infectious (I) at rate $\beta$ when in contact with an infectious node.
Infectious nodes recover spontaneously at rate $\mu$.
In the SIR model, nodes then enter an immune recovered (R) state, while in the SIS model, nodes become susceptible again and can be reinfected.
The quantities of interest are for the SIR model the distribution of epidemic sizes, defined as the final fraction of recovered nodes,
and for the SIS model the average fraction of infectious nodes $i_\infty$ in the stationary state.
We also calculate for the SIR model the fraction of epidemics that infect more than 20\% of the population and
the average size of these epidemics.
For the SIS model, we determine the epidemic threshold $\beta_{\rm c}$ for different values of $\mu$:
it corresponds to the value of $\beta$ that separates
an epidemic-free state ($i_\infty=0$) for $\beta<\beta_{\rm c}$ from an endemic state ($i_\infty > 0$) for $\beta>\beta_{\rm c}$, and
is thus an important indicator of the epidemic risk. We refer to the Methods section for further details on the  simulations.

We then present several methods for constructing surrogate data using only information contained in the resampled data. We
compare for each data set the outcomes of simulations performed on the whole data set,
on resampled data sets with a varying fraction of nodes removed, $f$, and
on the reconstructed data sets built using these various methods.

\subsection*{Uniformly resampled contact networks}

Missing data are known to affect the various properties of contact networks in different ways.
In particular, the number of neighbours (degree) of a node decreases as the fraction $f$ of removed nodes increases,
since removing nodes also removes links to these nodes. Under the hypothesis of uniform sampling, the average degree
$\langle k \rangle$  becomes $(1-f) \langle k \rangle$ for the resampled network \cite{Cohen:2000}. As a result,
the density of the resampled aggregated contact network, defined as the number of links divided by the total number of possible links between the nodes,
does not depend on $f$. The same reasoning applies to the density $\rho_{\rm AB}$ of links between groups of nodes $A$ and $B$, defined as the number of
links $E_{\rm AB}$ between nodes of group $A$ and nodes of group $B$, normalised by the maximum possible number of such links, $n_{\rm A} n_{\rm B}$,
where $n_{\rm A}$ is the number of nodes of group $A$ (for $A=B$, the maximum possible number of links is $n_{\rm A}(n_{\rm A} -1)/2.$):
both the expected number of neighbours of group $B$ for nodes of group $A$ (given by $E_{\rm AB}/n_{\rm A}$) and the number
$n_{\rm B}$ of nodes in group $B$ are indeed
reduced by a factor $(1-f)$, so that $\rho_{\rm AB}$ remains constant.
This means that the link density contact matrix, which gathers these densities and gives a measure of the interaction between
groups (here classes or departments), is stable under uniform resampling.
We illustrate these results on our empirical data sets in supplementary figures 1, 2, 4 and 5.
Table \ref{tab:sim_CM} and supplementary figure 2 show in particular that the similarities
between the original and resampled matrices are high for all data sets (see
supplementary figures 4--5 for the contact matrices themselves).

Finally, the temporal statistics of the contact network are not affected by population
sampling, as already noted in \cite{Cattuto:2010} for other data sets:
the distributions of contact and inter-contact durations (the inter-contact durations are the times
between consecutive contacts on a link), of
number of contacts per link and of cumulated contact durations (i.e.,
of the link weights in the aggregated network) do not change when the
network is sampled uniformly (supplementary figure 1). In the case of structured population,
an interesting property is moreover illustrated in supplementary figures 6--7:
although the distributions of contact durations occurring between members of the same group
or between individuals belonging to different groups are indistinguishable, this is not the
case for the distributions of the numbers of contacts per link nor, as a consequence, for
the distributions of cumulated contact durations. In fact, both cumulated contact durations and
numbers of contacts per link are more broadly distributed for links joining members of the same group.
The figures show that this property is stable under uniform resampling. 

\begin{figure}[ht]
  \includegraphics[width=\textwidth]{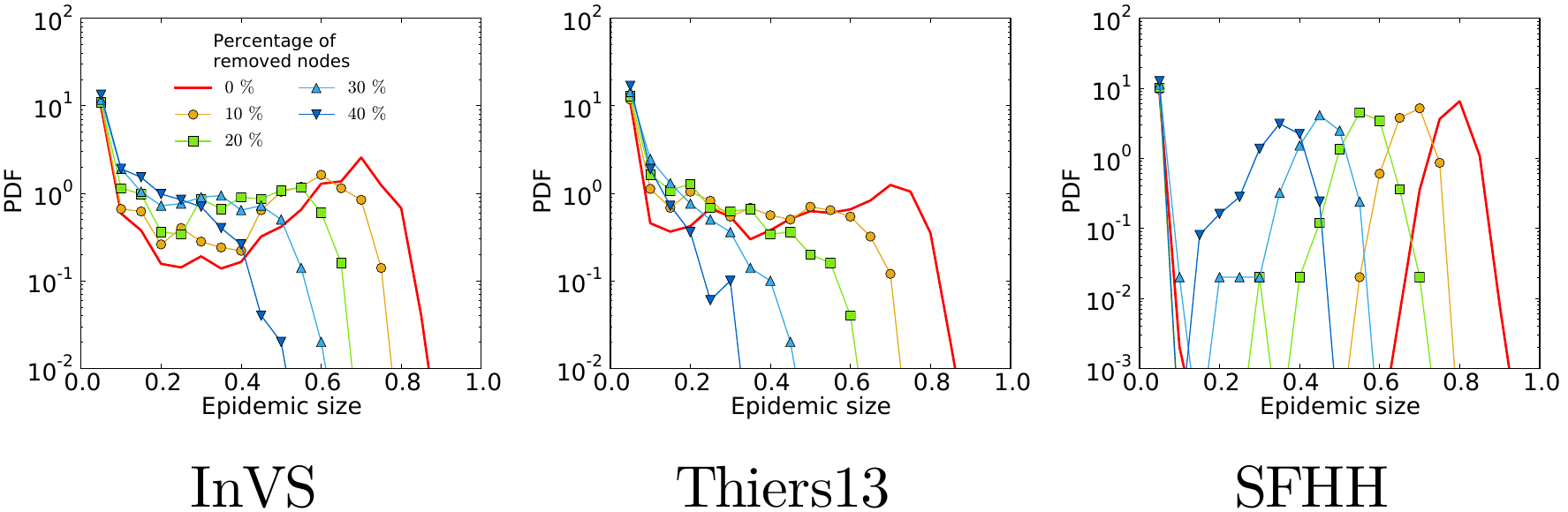}
  \caption{\label{sampling}{\bf SIR epidemic simulations on resampled contact networks.}
    We plot the distributions of epidemic sizes (fraction of recovered individuals) at the end of SIR processes simulated on top of resampled contact networks,
    for different values of the fraction $f$ of nodes removed. The plot shows the progressive disparition of large epidemic outbreaks as $f$ increases.
    The parameters of the SIR models are $\beta = 0.0004$ and $\beta/\mu = 1000$ (InVS) or $\beta/\mu = 100$ (Thiers13 and SFHH).
    The case $f=0$ corresponds to simulations using the whole data set, i.e., the reference case.
    For each value of $f$, $1,000$  independent simulations were performed.}
\end{figure}

Despite the robustness of these properties, the outcome
of simulations of epidemic spread is strongly affected by the
resampling. As  Fig.~\ref{sampling} illustrates for instance,
the probability of large outbreaks in the SIR model decreases strongly as $f$ increases and
even vanishes at  large  $f$. As mentioned above, such a result is expected, since the removed nodes act as if they
were immunised: sampling hinders the propagation in simulations by removing
transmission routes between the remaining nodes. As a consequence, the
prevalence and the final size of the outbreaks are systematically
underestimated by simulations of the SIR model on the resampled network with respect to
simulations on the whole data set (for the SIS model, the epidemic threshold is overestimated):
resampling leads overall to a systematic underestimation of the epidemic risk, and Fig. \ref{sampling}
illustrates the extent of this underestimation for the data at hand.

\subsection*{Estimation of epidemic sizes through simulations on reconstructed temporal networks}

We now present a series of methods to improve the estimation of the epidemic risk in simulations of epidemic spread
on temporal network data sets in which  nodes (individuals) are missing uniformly at random.
Note that we do not address here the problem of link prediction \cite{Liben-Nowell:2007} as our aim is not to infer the missing contacts.
The hierarchy of methods we put forward uses increasing amounts of information corresponding to increasing
amounts of detail on the group and temporal structure of the contact patterns, as measured in the resampled network.
We moreover assume that the timelines of scheduled activity are known (i.e., nights and weekends, during which no contact occurs).

For each data set, considered as ground truth, we create resampled data sets by removing at random a fraction $f$ of the $N$ nodes.
We then measure on each resampled data set a series of statistics of the resulting contact network and
construct stochastic, surrogate versions of the missing part of the network by
creating for each missing node a surrogate instance of its links and a synthetic timeline of contacts on each surrogate link, in the
different ways described below (see Supplementary Information and Methods section for more details on their practical implementation).

Method 0. As discussed above, the first effect of missing data is to decrease the average degree of the aggregate
contact network, while keeping its density constant. Hence, the simplest approach is to merely compensate this decrease. We therefore
measure the density of the resampled contact network $\rho_{\rm s}$, as well as the average aggregate duration of the contacts,
 $\langle w \rangle_{\rm s}$. We then add back
the missing nodes and create surrogate links between these nodes and between these nodes and the nodes of the resampled data set
at random, with the only constraint to keep the overall link density fixed to $\rho_{\rm s}$. We then attribute to each surrogate link the same
weight $\langle w \rangle_{\rm s}$ and create for each link a timeline of randomly chosen contact events of equal length $\Delta t = 20s$ (the temporal
resolution of the data set) whose total duration gives back $\langle w \rangle_{\rm s}$.

Method W. The heterogeneity of aggregated contact durations is known to play a role in the
spreading patterns of model diseases \cite{Read:2008,Smieszek:2009,Stehle:2011,Toth:2015}. We therefore refine Method 0
by collecting in the resampled data the list $\{w\}$ of aggregate contact durations, or weights (W). We build the surrogate links
and surrogate timelines of contacts on each link as in Method 0, except that each surrogate link carries
a weight extracted at random from $\{w\}$, instead of the average $\langle w \rangle_{\rm s}$.

Method WS. The fact that the population is divided into groups of individuals such as classes or departments can
have a strong impact on the structure of the contact network \cite{Stehle:2011a,Toth:2015} and on spreading processes \cite{Onnela:2007}.
We thus measure here the link density contact matrix of the resampled data, and
construct surrogate links in a way to keep this matrix fixed (equal to the value measured in the resampled data), in the
spirit of stochastic block models with fixed numbers of edges between blocks \cite{Peixoto:2012}. Moreover, we collect
in the resampled data two separate lists of aggregate contact durations: $\{w\}^{\rm int}$ gathers the weights of
links between individuals belonging to the same group, and $\{w\}^{\rm ext}$ is built with the weights of links joining individuals
of different groups. For each surrogate link, its weight is extracted at random either from $\{w\}^{\rm int}$ if it joins individuals
of the same group or from $\{w\}^{\rm ext}$ if it associates individuals of different groups. Timelines are then
attributed to links as in W. This method assumes that the number of missing nodes in each group is known,
and preserves the group structure (S) of the population.

Method WT. Several works have investigated how the temporal characteristics of networks (such as burstiness)
can slow down or accelerate spreading \cite{Karsai:2011,Holme:2012,Masuda:2013}. In order to take these
characteristics into account, we measure in the resampled data the
distributions of number of contacts per link and of contact and inter-contact durations, in addition to the global network density.
We build surrogate links as in Method W, and construct on each link a synthetic timeline in a way to respect the measured
temporal statistics (T) of contacts. More precisely, we attribute at random a number of contacts (taken
from the measured distribution) to each surrogate link, and then alternate contact and
inter-contact durations taken at random from the respective empirical distributions.

Method WST.  This method conserves the distribution of link weights (W), the group structure
(S), and the temporal characteristics of  contacts (T): surrogate links are built and weights assigned
as in method WS, and contact timelines on each link as in method WT.

Each of these methods uses a different amount of information gathered from the resampled data.
Methods 0, W and WT include an increasing amount of detail on the temporal structure of contacts:
method 0 assumes homogeneity of aggregated contact durations, while W takes into account their heterogeneity, and WT
reproduces heterogeneities of contact and inter-contact durations.
On the other hand, neither of these three methods assume any knowledge of the population
group structure. This can be due either to an effective complete lack of knowledge about the population structure, as in the SFHH data, or
also to the lack of data on the repartition of the missing nodes in the groups. Methods WS and WST on the other hand
reproduce the group structure as in a stochastic block model with fixed number of links within and between groups, and
take into account the difference between the distributions
of numbers of contacts and aggregate durations between individuals of the same or of different groups.
Indeed, links within groups correspond on average
to larger weights, as found empirically in \cite{Onnela:2007} and discussed above
(supplementary figures 5--6).
Overall, method WST is the one that uses most information measured in the resampled
data. (Additional properties such as the transitivity -which is also stable under resampling procedure,
see supplementary figure 3-
can also be measured in the resampled data and imposed
in the construction of surrogate links, as detailed
in the Supplementary Information. This comes however at a strong computational cost and we have verified
that it does not impact significantly our results, as shown in the supplementary figure 20.)

We check in Table~\ref{tab:sim_CM} and supplementary figures 8--13 that
the statistical properties of the resulting reconstructed (surrogate) networks, obtained by the union of the resampled data and
of the surrogate links, are similar to the ones of the original data for the WST method.
We emphasise again that our aim is not to infer the true missing contacts,
so that we do not compare the detailed structures of the surrogate and original contact networks.

Figures~\ref{fig:inferInVS}, \ref{fig:inferThiers}, \ref{fig:inferSFHH} and supplementary figures 16--19
display the outcome of SIR spreading simulations performed on surrogate networks obtained using the various reconstruction methods,
compared with the outcome of simulations on the resampled data sets, for various values of $f$.
Method 0 leads to a clear overestimation of the outcome and does not capture well the shape of the
distribution of outbreak sizes. Method W gives only slightly better results. The overall shape of the distribution
is  better captured for the three reconstruction methods using more information: WS, WT and WST
(note that for the SFHH case the population is not structured, so that W and WS are equivalent, as are
WT and WST). The WST method matches best the shape of the distributions and 
yields distributions much more similar to those obtained by simulating on the whole data set 
than the simulations performed on the resampled networks.
We also show in Fig.~\ref{fig:dist}
the fraction of outbreaks that reach at least $20\,\%$ of the population and the average epidemic size for these outbreaks.
In the case of simulations performed on resampled data, we rapidly lose information about the size and even the existence of large outbreaks
as $f$ increases. Simulations using data reconstructed with methods 0 and W,
on the contrary, largely overestimate these quantities, which is expected as
infections spread easier on random graphs than on structured graphs \cite{Onnela:2007,Karsai:2011}, especially
if the heterogeneity of the aggregated contact durations is not considered \cite{Stehle:2011,Toth:2015}.
Taking into account the population structure or using contact sequences that respect the temporal heterogeneities
(broad distributions of contact and inter-contact durations) yield better results
(WS and WT cases, respectively).
Overall, the WST method, for which the surrogate networks respect
all these constraints, yields the best results.

We show in the Supplementary Information that similar results are obtained for different values of the spreading parameters.
Moreover, as shown in Fig.~\ref{fig:SIS_InVS} and supplementary figures 14--15,
the phase diagram obtained for the SIS model when using reconstructed networks is much closer to the original
than for resampled networks. 
Overall, simulations on networks reconstructed using the WST method yield a much better estimation of the epidemic risk than simulations using
resampled network data, for both SIS and SIR models.

\subsection*{Reshuffled data sets.}

\begin{figure}[ht]
  \includegraphics[width=0.9\textwidth]{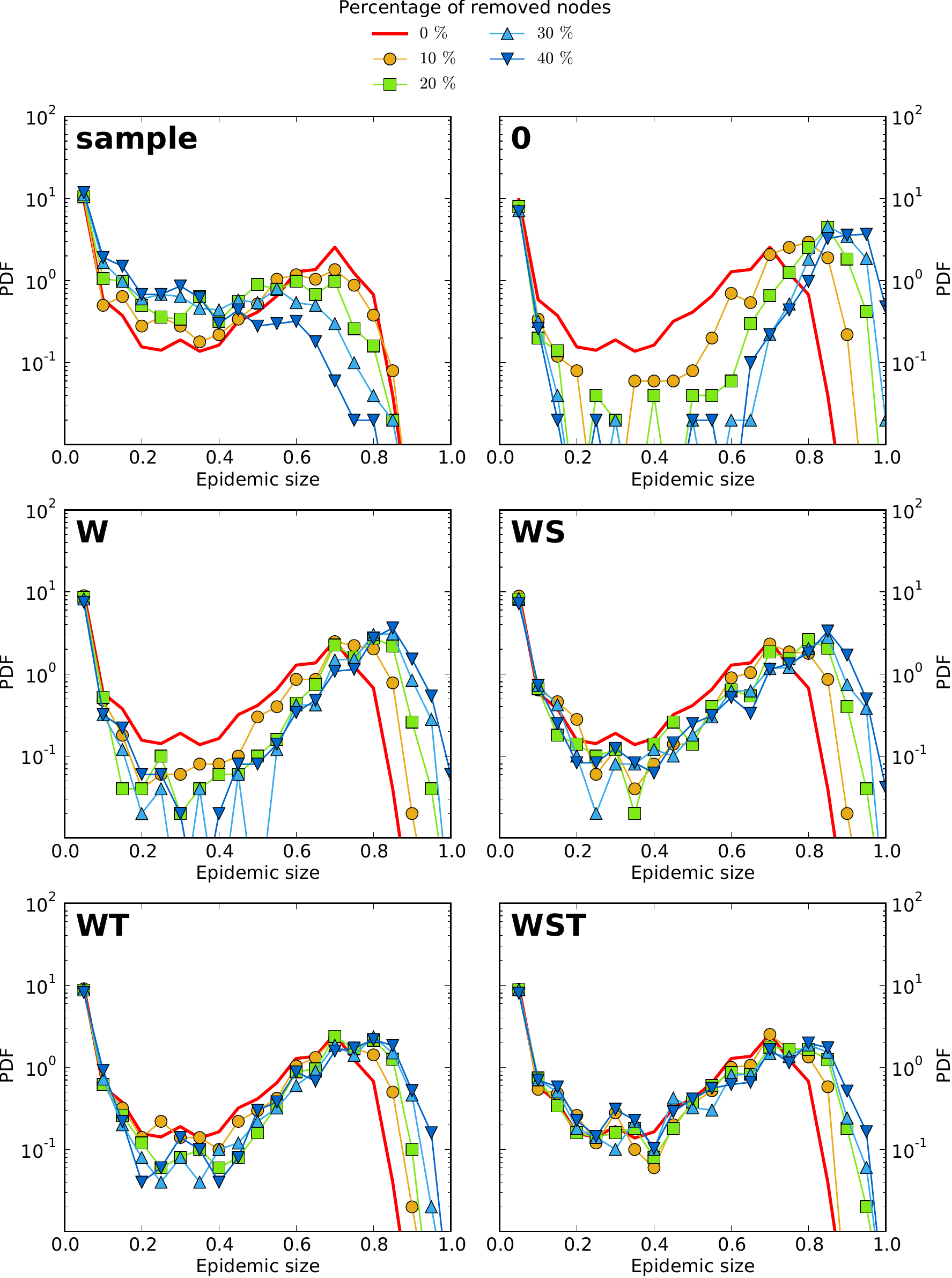}
  \caption{\label{fig:inferInVS}{\bf SIR simulations for the InVS (workplace) case.}
    We compare of the outcome of SIR epidemic simulations performed
    on resampled and reconstructed contact networks, for different methods of reconstruction.
    We plot the distribution of epidemic sizes (fraction of recovered individuals) at the end of SIR processes simulated on top of resampled (sample)
    and reconstructed contact networks, for different values of the fraction $f$ of nodes removed,
    and for the 5 reconstruction methods described in the text (0, W, WS, WT, WST).
    The parameters of the SIR models are $\beta = 0.0004$ and $\beta/\mu = 1000$.
    The case $f=0$ corresponds to simulations using the whole data set, \emph{i.e.}, the reference case.
    For each value of $f$, $1,000$ independent simulations were performed.}
\end{figure}

\begin{figure}[ht]
  \includegraphics[width=0.9\textwidth]{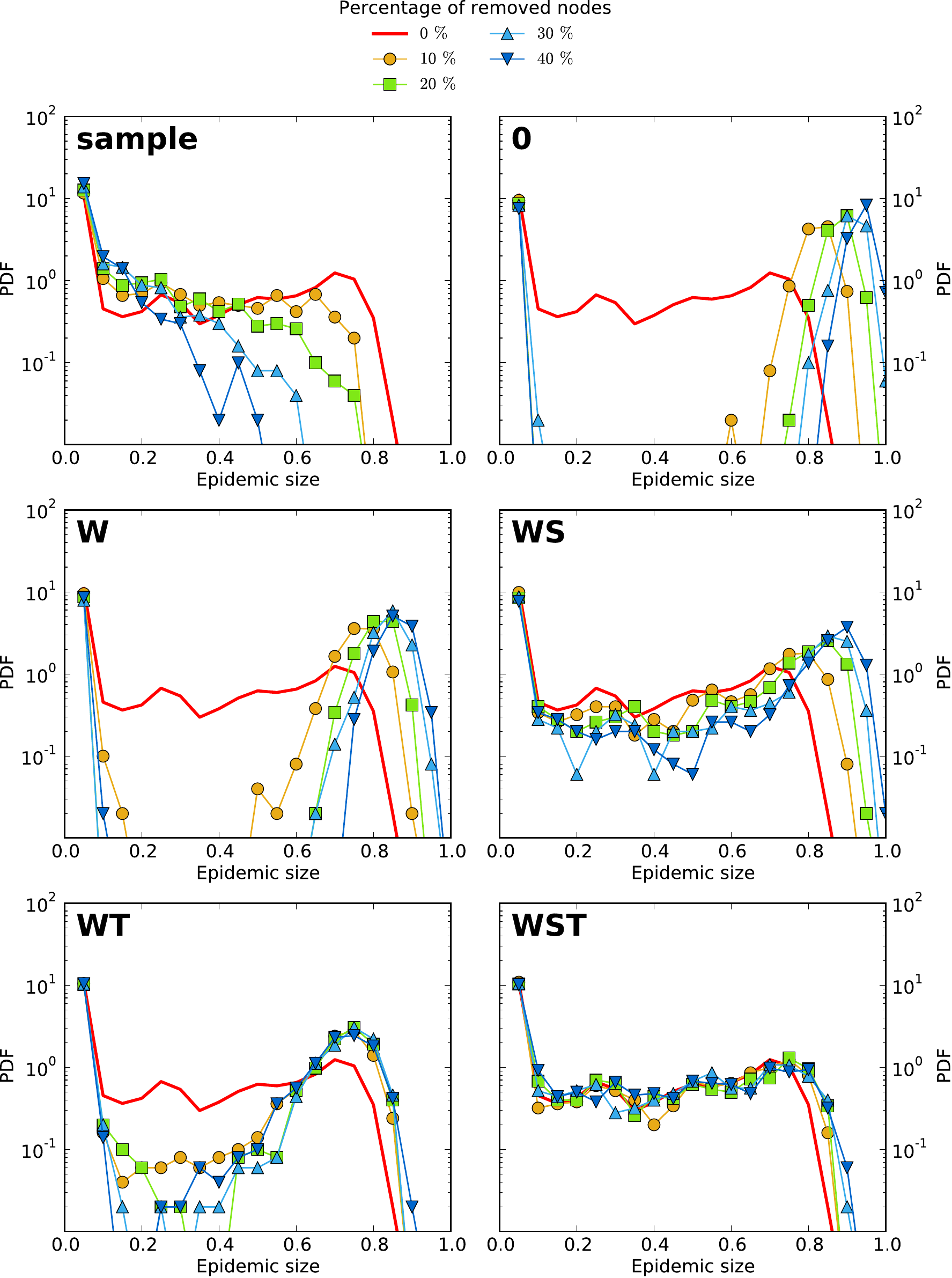}
  \caption{\label{fig:inferThiers}{\bf SIR simulations for the Thiers13 (high school) case.}
    We compare of the outcome of SIR epidemic simulations performed
    on resampled (top left) and  reconstructed contact networks, for different methods of reconstruction.
    We plot the distribution of epidemic sizes (fraction of recovered individuals) at the end of SIR processes simulated on top of resampled (sample)
    and reconstructed contact networks, for different values of the fraction $f$ of nodes removed,
    and for the 5 reconstruction methods described in the text (0, W, WS, WT, WST).
    The parameters of the SIR models are $\beta = 0.0004$ and $\beta/\mu = 100$.
    The case $f=0$ corresponds to simulations using the whole data set, i.e., the reference case.
    For each value of $f$, $1,000$ independent simulations were performed.}
\end{figure}

\begin{figure}[ht]
  \includegraphics[width=0.9\textwidth]{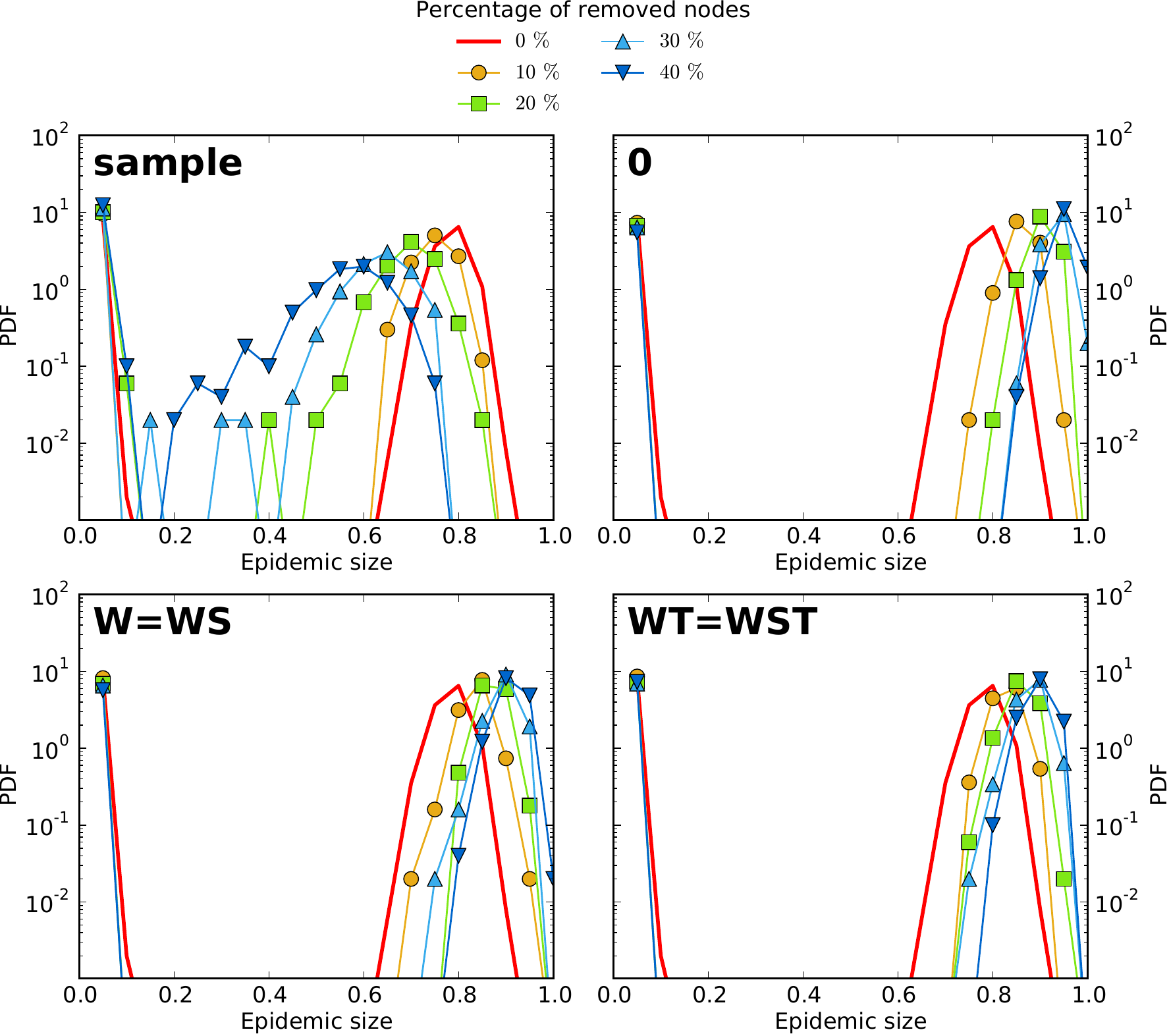}
  \caption{\label{fig:inferSFHH}{\bf SIR simulations for the SFHH (conference) case.}
    We compare of the outcome of SIR epidemic simulations performed
    on resampled and reconstructed contact networks, for different methods of reconstruction.
    We plot the distribution of epidemic sizes (fraction of recovered individuals) at the end of SIR processes simulated on top of resampled (sample)
    and reconstructed contact networks, for different values of the fraction $f$ of nodes removed,
    and for three reconstruction methods described in the text (0, W, WT).
    In this case, as the population is not structured, methods W and WS on the one hand, WT and WST on the other hand, are equivalent.
    The parameters of the SIR models are $\beta = 0.0004$ and $\beta/\mu = 100$.
    The case $f=0$ corresponds to simulations using the whole data set, i.e., the reference case.
    For each value of $f$, $1,000$ independent simulations were performed.}
\end{figure}

Even when simulations are performed on reconstructed contact patterns built with the WST method, the
maximal outbreak sizes are systematically overestimated (Figs.~\ref{fig:inferInVS} - \ref{fig:inferSFHH}),
as well as, in most cases, the probability and average size of large outbreaks,
especially for the SFHH case (Figs. \ref{fig:inferSFHH} - \ref{fig:dist}).
These discrepancies might stem from structural and/or temporal correlations present in
empirical contact data that are not taken into account in our reconstruction methods.
In order to test this hypothesis, we construct several reshuffled data sets and use them as initial data in our resampling and
reconstruction procedure. We use both structural and temporal reshuffling as described in the Methods section, in order
to remove either structural correlations, temporal correlations, or both, from the original data sets. We then proceed to a resampling and reconstruction
procedure (using the WST method) as for the original data, and perform numerical simulations of SIR processes.
As for the original data, simulations on resampled data lead to a strong underestimation of the process outcome, and
simulations using the reconstructed data gives much better results.

We show in the supplementary figures 21-22 that we still obtain discrepancies, and in particular
overestimations of the largest epidemic sizes,  when we use 
temporally reshuffled data in which the link structure of the contact network is maintained. If on the other hand 
we use data in which the network structure has been reshuffled in a way to cancel structural correlations within each group, 
the reconstruction procedure gives a very good agreement
between the distributions of epidemic sizes of original and reconstructed data, as shown in
Fig. \ref{fig:inferCMshuffled}.  More precisely we consider
here ``CM-shuffled'' data, i.e., contact networks in which the links have been reshuffled randomly but separately for each pair of groups, i.e., a link between an
individual of group $A$ and an individual of group $B$ is still between groups $A$ and $B$ in the reshuffled network.
The difference with the case of non-reshuffled empirical data 
is particularly clear for the SFHH case. This  indicates that
the overestimation observed in Figs. \ref{fig:inferInVS} - \ref{fig:inferSFHH} is
mostly due to the fact that the reconstructed data does not reproduce small scale structures
of the contact networks: such structures might be due to e.g. groups of colleagues or friends, whose composition is neither available as metadata nor
detectable in the resampled data sets.

\subsection*{Limitations.}

\begin{figure}[ht]
  \includegraphics[width=\textwidth]{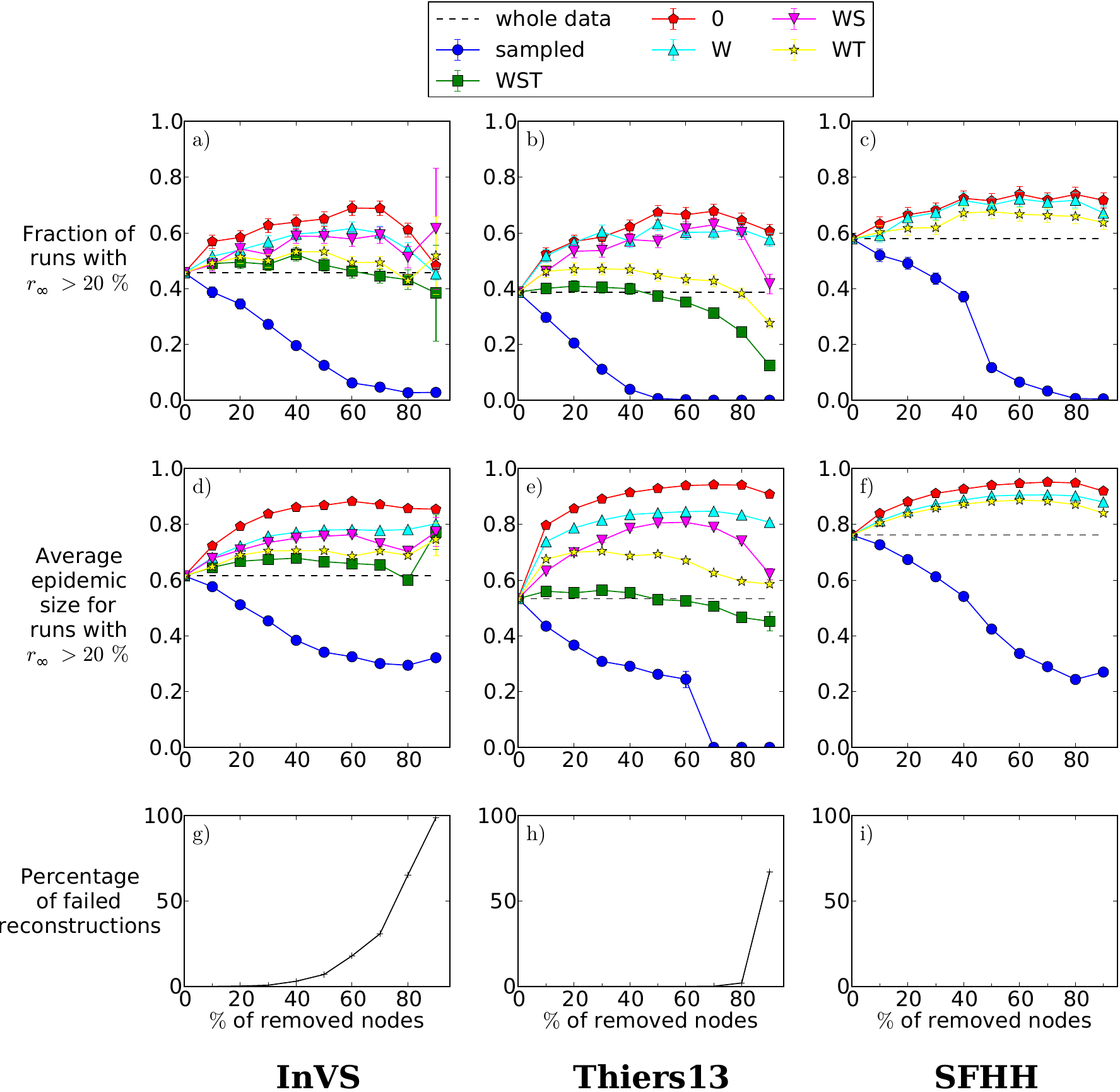}
  \caption{\label{fig:dist}{\bf Accuracy of the different reconstruction methods.}
    We perform SIR epidemic simulations for each case, for different values of the fraction $f$ of missing nodes,
    for both sampled networks and networks reconstructed with the different methods.
    We compare in each case, and as a function of $f$, the fraction of outbreaks that lead to a
    final fraction of recovered individuals $r_\infty$ larger than $20\,\%$ of the population (a, b, c),
    and the average size of these large outbreaks (d, e, f).
    The dashed lines give the corresponding values for simulations performed on the complete data sets.
    The different methods are:
    reconstruction conserving only the link density and the average weight of the resampled data (0);
    reconstruction conserving only the link density and the distribution of weights of the resampled data (W);
    reconstruction preserving, in addition to the W method, the group structure of the resampled data (WS);
    reconstruction conserving link density, distribution of weights and distributions of contact times,
    of inter-contact times and of numbers of contacts per link measured in the resampled data (WT);
    full method conserving all these properties (WST).
    We also plot as a function of $f$ the failure rate of the WST algorithm,
    \emph{i.e.}, the percentage of failed reconstructions (g, h, i).
    For the SFHH case, as the population is not structured into groups, methods W and WS are equivalent, as well as methods WT and WST;
    moreover, reconstruction is always possible.
    The SIR parameters are $\beta = 0.0004$ and $\beta/\mu = 1000$ (InVS) or $\beta/\mu = 100$ (Thiers13 and SFHH)
    and each point is averaged over $1,000$ independent simulations.}
\end{figure}

\begin{figure}[ht]
  \includegraphics[width=\textwidth]{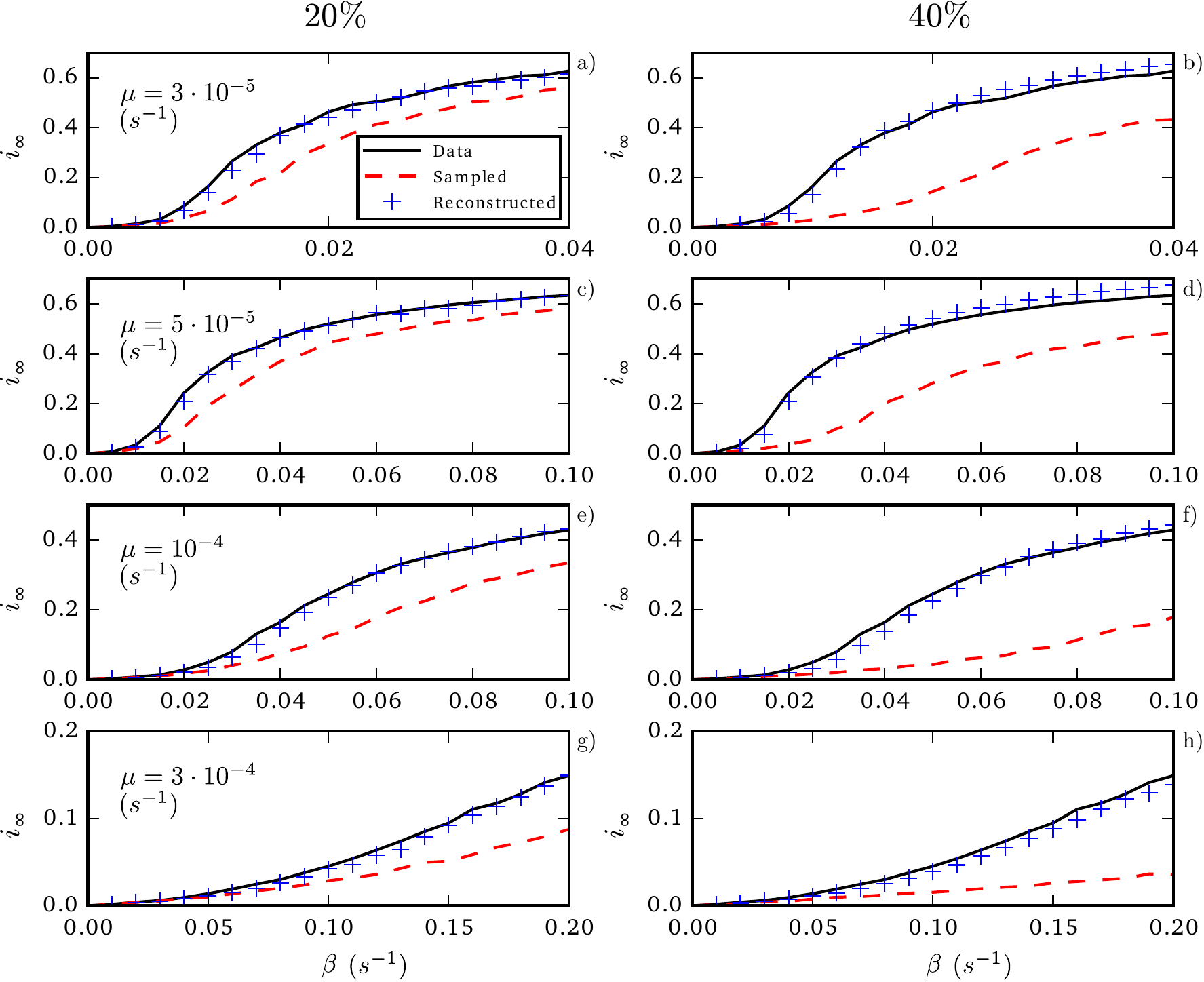}
  \caption{\label{fig:SIS_InVS}{\bf SIS simulations for the InVS (workplace) case.}
    We perform SIS epidemic simulations and report the phase diagram of the SIS model
    for the original, resampled and reconstructed contact networks.
    Each panel shows the stationary value $i_\infty$ of the prevalence in the stationary state of the SIS model,
    computed as described in Methods, as a function of $\beta$, for several values of $\mu$.
    Simulations are performed in each case using either the complete data set (continuous lines),
    resampled data (dashed lines) or reconstructed contact networks using the WST method (pluses).
    The fraction of excluded nodes in the resampling is $f=20\%$ for a, c, e, g and  $f=40\%$ for b, d, f, h.
}
\end{figure}

\begin{figure}[ht]
  \includegraphics[width=0.9\textwidth]{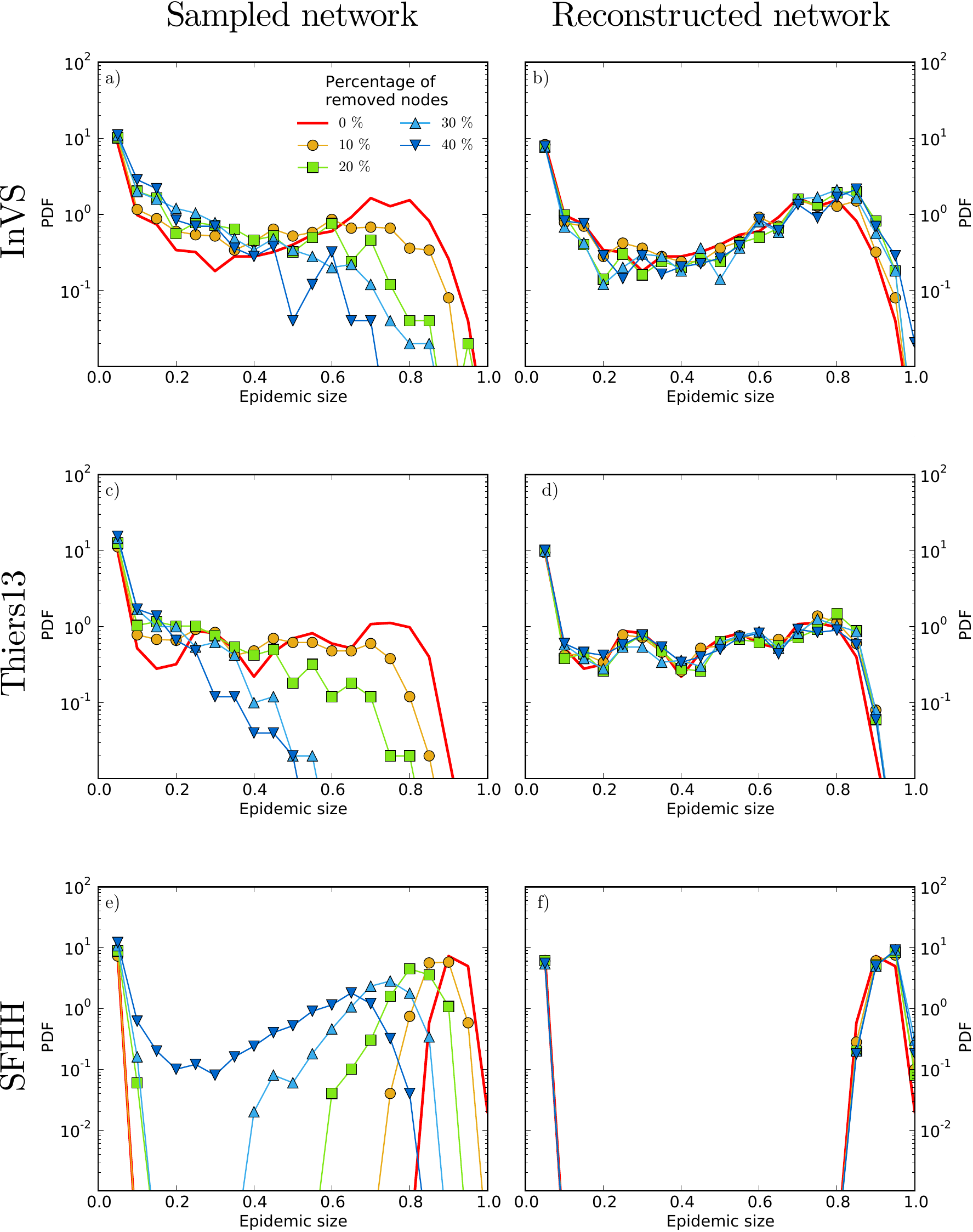}
  \caption{\label{fig:inferCMshuffled}{\bf SIR simulations on shuffled data.}
    We compare of the outcome of SIR epidemic simulations performed on resampled and reconstructed contact networks, for shuffled data.
    We plot the distribution of epidemic sizes (fraction of recovered individuals) at the end of SIR processes simulated
    on top of either resampled (a, c, e) or reconstructed  (b, d, f) contact networks,
    for different values of the fraction $f$ of nodes removed.
    We use here the WST reconstruction method, and the data set considered consists in a CM-shuffled version (see Methods)
    of the original data, in which the shuffling procedure removes structural correlations of the contact network within each group.
    The parameters of the SIR models are $\beta = 0.0004$ and $\beta/\mu = 1000$ (InVS) or $\beta/\mu = 100$ (Thiers13 and SFHH).
    The case $f=0$ corresponds to simulations using the whole data set, \emph{i.e.}, the reference (reshuffled data) case.
    For each value of $f$, $1,000$  independent simulations were performed.}
\end{figure}

\begin{figure}[ht]
  \includegraphics[width=\textwidth]{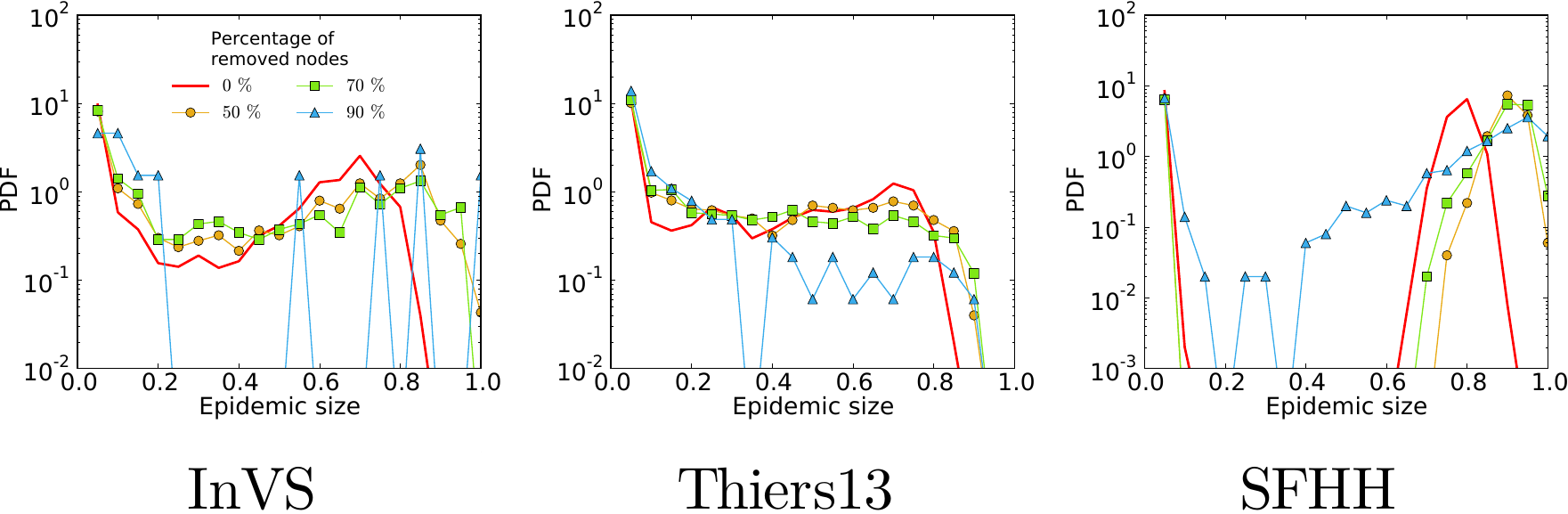}
  \caption{\label{fig:infer_high}{\bf SIR simulations for very large numbers of missing nodes.}
    We simulate SIR processes on reconstructed contact networks for large values of the fraction $f$ of removed nodes.
    We plot the distributions of epidemic sizes for simulations on reconstructed networks and on the whole data set (case $f=0$),
    for large values of the fraction $f$ of removed nodes.
    Here $\beta = 0.0004$ and $\beta/\mu = 1000$ (InVS) or $\beta/\mu = 100$ (Thiers13 and SFHH) and $1,000$ simulations were performed for each value of $f$.
    The distributions of epidemic sizes for simulations performed on resampled data sets are not shown since at these high values of $f$,
    almost no epidemics occur.}
\end{figure}

When the fraction $f$ of nodes excluded by the resampling procedure becomes large, the properties
of the resampled data may start to differ substantially from those of the whole data set (Figs. S1 \& S2).
As a result,  the distributions of epidemic sizes of SIR simulations show stronger deviations from
those obtained on the whole data set (Fig.~\ref{fig:infer_high}), even if the epidemic risk evaluation is still better than
for simulations on the resampled networks (Fig. \ref{fig:dist}).
Most importantly however, the information remaining in the resampled data at large $f$ can be insufficient
to construct surrogate contacts. This happens in particular if an entire class or department is
absent from the resampled data or if all the resampled nodes of a class/department are disconnected (see Methods
for details).  We show in the bottom plots of Fig.~\ref{fig:dist}
the failure rate, i.e., the fraction of cases in which we are not able to construct
surrogate networks from the resampled data. The failure rate increases gradually with $f$ for the InVS data since
the groups (departments) are of different sizes. For the Thiers13 data, all classes
are of similar sizes so that the failure rate reaches abruptly a large value at a given value of $f$.
For the SFHH data, we can always construct surrogate networks as the population is not structured.
Another limitation of the reconstruction method lies in the need to know the number of individuals missing
in each department or class. If these numbers are completely unknown, giving an estimation of outbreak sizes is impossible
as adding arbitrary numbers of nodes and links to the resampled data can lead to arbitrarily large epidemics.
The methods are however still usable if only partial information is available. For instance, if
only the overall missing number of individuals is available, it is possible to use
the WT method, which still gives sensible results. Moreover, if $f$ is only approximately known, e.g., $f$ is known
to be within an interval of possible values $[f_1,f_2]$, it is possible to perform two reconstructions using the respective hypothesis
$f=f_1$ and $f=f_2$ and to give an interval of estimates. We provide an example of such procedure in supplementary figure 23.

\section*{Discussion}

The understanding of epidemic spreading phenomena has been vastly improved thanks to the use of data-driven models
at different scales. High resolution contact data in particular
have been used to evaluate epidemic risk or containment policies in specific populations 
or to perform contact tracing
\cite{Salathe:2010,Smieszek:2013,Chowell:2013,Gemmetto:2014,Voirin:2015,Obadia:2015,Toth:2015}. In such studies, missing data due to
population sampling might represent however
a serious issue: individuals absent from a data set are indeed equivalent to immunised individuals when epidemic processes are simulated.
Feeding sampled data into data-driven models can therefore lead to severe underestimations of the epidemic risk
and might even a priori affect the  evaluation of mitigation strategies if for instance
some at-risk groups are particularly undersampled.

Here we have put forward a set of methods to obtain a better evaluation of the outcome of spreading simulations for
data-driven models using contact data from a uniformly sampled population.
To this aim, we have shown how it is possible, starting from a data set describing the contacts of only a fraction of the population of interest
(uniformly sampled from the whole population),
to construct surrogate data sets using various amounts of accessible information, i.e., quantities measured in the sampled data.
We have shown that the simplest method, which consists in simply compensating for the decrease in the average number of neighbours due
to sampling, yields a strong overestimation of the epidemic risk. When additional information describing
the group structure and the temporal properties of the data
is added in the construction of surrogate data sets,
simulations of epidemic spreading on such surrogate data yield results similar to those obtained on the complete data set.
(We note that the  issue of how much information should be included when constructing the surrogate data
is linked to the general issue of how much information is needed to get an accurate picture of spreading processes on temporal networks
\cite{Stehle:2011,Blower:2011,Machens:2013,Smieszek:2013,Chowell:2013,Pfitzner:2013}.)
Some discrepancies in the epidemic risk estimation are however still observed, due in particular to
small scale structural correlations of the contact network that are difficult or even impossible to measure in the resampled data: 
these discrepancies are indeed largely suppressed
if we use as original data a reshuffled contact network in which such correlations are absent.

The methods presented here yield much better results than simulations using resampled data,
even when a substantial part of the population is excluded,
in particular in estimating the probability of large outbreaks.
It suffers however from limitations, especially when the fraction $f$ of excluded individuals is too large.
First, the construction of the surrogate contacts relies on the stability of a set of quantities with respect to resampling, but
the measured quantities start to deviate from the original ones at large $f$.
The shape of the distribution of epidemic sizes may then differ substantially from the original one.
Second, large values of $f$ might even render the construction of the surrogate data impossible due to the loss of information on whole
categories of nodes. Finally, at least an estimate of the number of missing individuals in the population is needed in order
to create a surrogate data set.

An interesting avenue for future work concerns possible improvements of the reconstruction methods, in particular
by integrating into the surrogate data additional information and complex correlation patterns
measured in the sampled data. For instance,
the number of contacts varies significantly with the time of day in most contexts:
the corresponding activity timeline
might be measured in the sampled data (overall or even for each group
of individuals), assumed to be robust to sampling and used in the reconstruction
of contact timelines.
More systematically, it might also be possible to use the
temporal network decomposition technique put forward in  \cite{Gauvin:2014} on the sampled data,
in order to extract mesostructures such as temporally-localized mixing patterns. The
surrogate contacts could then be built in a way to preserve such patterns.
Indeed, correlations between structure and activity in the temporal contact network
are known to influence spreading processes
 \cite{Karsai:2011,Isella:2011,Pfitzner:2013,Gauvin:2013,Scholtes:2014,Gauvin:2015}
but are notoriously difficult to measure.
If the group structure of the population is unknown, recent approaches based on stochastic block models \cite{Peixoto:2015}
might be used to extract groups from the resampled data; this extracted group structure could then be used to build the corresponding contact matrix
and surrogate data sets.

We finally recall that we have assumed an uniform sampling of nodes, corresponding to
an independent random choice of each individual of the population to take part or not to the data collection.
Other types of sampling or data losses can however also be present in data collected by wearable sensors, such as
partial coverage of the premises of interest by the measuring infrastructure, non-uniform sampling
depending on individual activity (too busy persons or, on the contrary, asocial individuals, might not want to wear sensors),
on group membership, or due to clusters of non-participating individuals (e.g., groups of friends).  In addition, other types of data sets such as the ones
obtained from surveys or diaries correspond to different types of sampling, as each respondent
provides then information in the form of an ego-network \cite{Robins:2004}. Such data sets
 involve potentially
additional types of biases such as underreporting of the number of contacts
and overestimation of contact durations \cite{Smieszek:2012,Smieszek:2014,Mastrandrea:2015}: how to adapt the methods presented
here is an important issue that we will examine in future work.
Finally, the population
under study is (usually) not isolated from the external world, and it would be important to devise ways to include contacts with outsiders
in the data and simulations, for instance by using other data sources such as surveys.

\section*{Acknowledgements}

The present work is partially
supported by the French ANR project HarMS-flu (ANR-12-MONU-0018) to M.G. and A.B,
by the EU FET project Multiplex 317532 to A.B., C.C. and C.L.V.,
by the A*MIDEX project (ANR-11-IDEX-0001-02) funded by the ``Investissements d'Avenir'' French Government program,
managed by the French National Research Agency (ANR) to A.B.,
by the Lagrange Project of the ISI Foundation funded by the CRT Foundation to C.C.,
and by the Q-ARACNE project funded by the Fondazione Compagnia di San Paolo to C.C.

\section*{Author contributions}
A.B. and C.C. designed and supervised the study.
M.G., C.L.V., C.C., and A.B. collected and post-processed the data,
analyzed the data, carried out computer simulations and prepared the figures.
M.G., C.L.V., C.C., and A.B. wrote the manuscript.

The authors declare that no competing financial interests exist.

\section*{Methods}

\subsection*{Data}

We consider data sets collected using the SocioPatterns proximity sensing platform (\url{http://www.sociopatterns.org}) based
on wearable sensors that detect close face-to-face proximity of individuals wearing them.
Informed consent was obtained from all participants and the French national bodies responsible for ethics and privacy,
 the Commission Nationale de l'Informatique et des
Libert\'es (CNIL, http://www.cnil.fr), approved the data collections.

The high school (Thiers13) data set  \cite{Mastrandrea:2015} is structured in 9 classes, forming three subgroups
of three classes corresponding to their specialisation in
Mathematics-Physics (MP, MP$^*$1, MP$^*$2 with respectively $31$, $29$ and $38$ students),
Physics (PC,  PC$^*$, PSI with respectively $44$, $39$ and $34$ students),
or Biology (2BIO1, 2BIO2, 2BIO3  with respectively $37$, $35$ and $39$ students).

The workplace (InVS) data set \cite{Genois:2015} is structured in $5$ departments:
DISQ (Scientific Direction, $15$ persons), DMCT (Department of Chronic Diseases and Traumatisms, $26$ persons),
DSE (Department of Health and Environment, $34$ persons), SRH (Human Resources, $13$ persons) and SFLE (Logistics, $4$ persons).

For the conference data (SFHH), we
do not have metadata on the participants, and the aggregated network structure was
found to be homogeneous \cite{Stehle:2011}.

\subsection*{SIR and SIS simulations}

Simulations of SIR and SIS processes on the temporal networks of contacts (original, resampled or reconstructed) are performed using the temporal 
Gillespie algorithm described in~\cite{Vestergaard2015}.
For each run of the simulations, all nodes are initially susceptible; a node is chosen at random as the seed of the epidemic and 
put in the infectious state at a point in time chosen at random over the duration of the contact data.
A susceptible node in contact with an infectious node becomes infectious at rate $\beta$.
Infectious nodes recover at rate $\mu$: in the SIR model they then enter the recovered state and cannot become infectious again, 
while in the SIS model they enter the susceptible state again.
If needed, the sequence of contacts is repeated in the simulation~\cite{Stehle:2011}.

For SIR processes, we run each simulation, with the seed node chosen at random, until no infectious individual remains (nodes are thus either still susceptible or have been infected and then recovered).
We consider values of $\beta$ and $\mu$ yielding a non-negligible epidemic risk, i.e.,
such that a rather large fraction of simulations lead to a final size larger than $20\%$ of the population (see Figs. \ref{sampling}--\ref{fig:inferSFHH}):
$\beta = 4\times 10^{-4} s^{-1}$, $\mu = 4\times 10^{-7} s^{-1}$ (InVS) or $4\times 10^{-6} s^{-1}$ (SFHH and Thiers13).
Other parameter values are explored in the Supplementary Information.
For each set of parameters, the distribution of epidemic sizes is obtained by performing $1,000$ simulations.

For SIS processes, simulations are performed using the quasi-stationary approach of~\cite{Ferreira2011}. They are run until the system enters a stationary state as witnessed by the mean number of infected nodes being constant over time. Simulations are then continued for 50,000 time-steps while recording the number of infected nodes. For each set of parameters, the simulations are performed once with each node of the network chosen as the seed node.

\subsection*{Reconstruction algorithm}

We consider a population ${\cal P}$ of $N$ individuals, potentially organised in groups.
We assume that all the contacts occurring among a subpopulation ${\cal \tilde{P}}$ of these individuals,
of size $\tilde{N} = (1-f) N$, are known. This constitutes our resampled data from which we
need to construct a surrogate set of contacts concerning the remaining $n = N - \tilde{N} = f N$ individuals 
for which no contact information is available: these contacts can occur among these individuals and between them
and the members of ${\cal \tilde{P}}$. We assume that we know the group  of each member of ${\cal P} \backslash {\cal \tilde{P}}$, 
and the overall activity timeline, i.e., the intervals during which contacts take place, 
separated by nights and weekends.

To construct the surrogate data (WST method), we first compute from the activity timeline the total duration $T_{\rm u}$ of the periods during which contacts can occur.

Then, we measure in the sampled data:
\begin{itemize}
\item the density $\rho$ of links in the aggregated contact network;
\item a row-normalised contact matrix $C$, 
  in which the element $C_{\rm AB}$ gives the probability for a node of group $A$ to have a link to a node of group $B$;
\item the list $\{\tau_{\rm c}\}$ of contact durations;
\item the lists $\{\tau_{\rm ic}\}^{\rm int}$ and $\{\tau_{\rm ic}\}^{\rm ext}$ of inter-contact durations for internal and external links, 
  \emph{i.e.}, for links between nodes of the same group and links between nodes that belong to two different groups, respectively;
\item the lists $\{p\}^{\rm int}$ and $\{p\}^{\rm ext}$ of numbers of contacts per link, respectively for internal (within groups) and external (between groups)
  links;
\item the list $\{t_0\}$ of initial times between the start of the data set and the first contact between two nodes.
\end{itemize}

Given $\rho$, we compute the number $e$ of additional links needed to keep the network density constant when we add the $n$ excluded nodes.

We then construct each link according to the following procedure:
\begin{itemize}
\item a node $i$ is randomly chosen from the set ${\cal P} \backslash {\cal \tilde{P}}$ of excluded nodes;
\item knowing the group $A$ that $i$ belongs to, we extract at random a target group $B$ with probability given by $C_{\rm AB}$;
\item we draw a target node $j$ at random from $B$ (if $B=A$, we take care that $i\ne j$) such that $i$ and $j$ are not linked;
\item depending on whether $i$ and $j$ belong to the same group or not, we draw from $\{p\}^{\rm int}$ or $\{p\}^{\rm ext}$ the number of contact events $p$ taking place over the link $ij$;
\item from $\{t_0\}$, we draw the initial waiting time before the first contact;
\item from $\{\tau_{\rm c}\}$, we draw $p$ contact durations $\tau_{\rm c}^k$, $k=1,\cdots,p$;
\item from $\{\tau_{\rm ic}\}^{\rm int}$ or $\{\tau_{\rm ic}\}^{\rm ext}$, we draw $p-1$ inter-contact durations $\tau_{\rm ic}^m$, $m=1,\cdots,p-1$;
\item if $t_0 + \sum_k \tau_{\rm c}^k + \sum_m \tau_{\rm ic}^m > T_{\rm u}$, we repeat steps (d) to (g) until we obtain a set of values such that
  $t_0 + \sum_k \tau_{\rm c}^k + \sum_m \tau_{\rm ic}^m \le T_{\rm u}$;
\item from $t_0$ and the $\tau_{\rm c}^k$ and $\tau_{\rm ic}^m$, we build the contact timeline of the link $ij$;
\item finally, we insert in the contact timeline the breaks defined by the global activity timeline.
\end{itemize}

\subsection*{Possible failure of the reconstruction method at large $f$}

The construction of the surrogate version of the missing links uses as an input the group structure of the subgraph that remains after sampling, as
given by the contact matrix of the link densities between the different groups of nodes that are present in the subpopulation ${\cal \tilde{P}}$.
Depending on the characteristics of ${\cal \tilde{P}}$ and of the corresponding contacts, the construction method can fail in several cases:
(i) if an entire group (class/department) of nodes in the population is absent from ${\cal \tilde{P}}$;
(ii) if the remaining nodes of a specific group (class/department) are all isolated in ${\cal \tilde{P}}$'s contact network;
(iii) if, during the algorithm, a node of ${\cal P} \backslash {\cal \tilde{P}}$ is selected in a certain group $A$ but cannot create any more links
because it already has links to all nodes in the groups $B$ such that $C_{\rm AB} \ne 0$;
(iv) if there are either no internal (within groups) or external (between groups) 
links in the contact network of ${\cal \tilde{P}}$: in this case one of the lists of link temporal 
characteristics is empty and the corresponding structures cannot be reconstructed.

Cases (i) and (ii) correspond to a complete loss of information about the connectivity of a group (class/department) of the population, due to sampling. It is
then impossible to reconstruct a sensible connectivity pattern for these nodes.
Case (iii) is more subtle and occurs in situations of very low connectivity between groups. For instance, within the contact network of
${\cal P}$, a group $A$ has links only with another specific group $B$, and both $A$ and $B$ are small; it is then possible that 
the nodes of $({\cal P} \backslash {\cal \tilde{P}}) \cap A$ exhaust the set of possible links to nodes of $B$ during the reconstruction
algorithm. If a node of $({\cal P} \backslash {\cal \tilde{P}})  \cap A$ is again chosen to create a link, such a creation is not possible and 
the construction fails.
Case (iv) usually corresponds to situations in which the links between individuals of different groups which remain in the
resampled data set correspond to pairs of individuals who have had only 
one contact event: in such cases, 
$\{\tau_{\rm ic}\}^{\rm ext}$ is empty and external links with more than one contact cannot be built.

\subsection*{Shufflings}

In order to test the effect of correlations in the temporal network, we use four shuffling methods, based on the ones defined in \cite{Gauvin:2013}.

Link shuffling. The contact timelines associated with each link are randomly redistributed among the links. Correlations between 
timelines of links adjacent to a given node are destroyed, as well as correlations between weights and topology. The structure of the network is kept,
as well as the global activity timeline.

Time shuffling. From the contact data we build the lists $\{\tau_{\rm c}\}$, $\{\tau_{\rm ic}\}$ and $\{p\}$ of, respectively, 
contact durations, inter-contact durations and number of contacts per link. We also measure 
the list $\{t_0\}$ of initial times between the start of the data set and the first contact between two nodes.
For each link, we draw randomly a starting time $t_0$, a number $p$ of contacts from $\{p\}$, $p$ contact durations from 
$\{\tau_{\rm c}\}$ and $p-1$ inter-contact durations from $\{\tau_{\rm ic}\}$, so that the total duration of the timeline does not exceed 
the total available time $T_{\rm u}$. We then construct the contact timelines, thus destroying the temporal correlations among contacts.
The structure of the network is instead kept fixed.

CM shuffling. We perform a link rewiring separately on each 
compartment of the contact matrix, \emph{i.e.}, we randomly redistribute links with their contact timelines within each group, and within each pair of groups. 
We thus destroy the structural correlations inside each compartment of the contact matrix, while preserving the group structure of the network
as given by the link density contact matrix and the contact matrix of total contact times between groups.

CM-time shuffling. We perform both a CM shuffling and a time shuffling.

\onecolumngrid

\clearpage

\section*{Tables \& Figures}

\begin{table}[ht]
  \begin{tabular}{|c|c|c|c|c|c|}
    \hline
    Data set&Type&$N$&$r$&$T$& Dates\\
    \hline
    InVS&Workplace & 92 & 63\,\%&2 weeks&June 24th - July 5th 2013\\
    Thiers13&High school & 326 &  86\,\%&1 week& December 2nd - 7th 2013\\
    SFHH&Conference& 403 &  34\,\%&2\,days&June 3rd - 4th 2009\\
    \hline
  \end{tabular}
  \caption{\label{tab:data}\textbf{Data sets.}
    For each data set we specify the type of social situation, the number
    $N$ of individuals whose contacts were measured, the
    corresponding participation rate $r$, the duration $T$ and the dates of the data collection.}
\end{table}

\begin{table}[ht]
  \begin{tabular}{|c|c|c|c|c|c|}
    \hline
                  &$f$     &InVS CML              &Thiers13 CML\\
    \hline
    \hline
                  &$10\,\%$&$0.996\ [0.937,0.999]$& $0.999\ [0.998,0.999]$\\
    Resampled     &$20\,\%$&$0.980\ [0.889,0.994]$& $0.996\ [0.995,0.997]$\\
                  &$40\,\%$&$0.925\ [0.872,0.983]$& $0.988\ [0.983,0.990]$\\
    \hline
    \hline
                  &$10\,\%$&$0.976\ [0.846,0.995]$&  $0.998\ [0.994,0.999]$\\
    Reconstructed &$20\,\%$&$0.942\ [0.844,0.984]$&  $0.993\ [0.985,0.995]$\\
                  &$40\,\%$&$0.890\ [0.652,0.953]$&  $0.977\ [0.938,0.987]$\\
    \hline
  \end{tabular}
  \caption{\label{tab:sim_CM}\textbf{Contact matrix similarities}
    Similarities between the original contact matrices and the contact matrices of the resampled networks (top) and of the
    reconstructed networks (bottom). Median and 90\% confidence interval for the cosine similarity between link density
    contact matrices (CML) for different values of $f$, the fraction of nodes removed from the original data.
    Values were obtained from $100$ independent realisations of the resampling and reconstruction procedures.
  }
\end{table}


\clearpage

\renewcommand{\figurename}{Supplementary Figure} 


\section*{Supplementary figures}

 \begin{figure*}[htp]
   \includegraphics[width=0.95\textwidth]{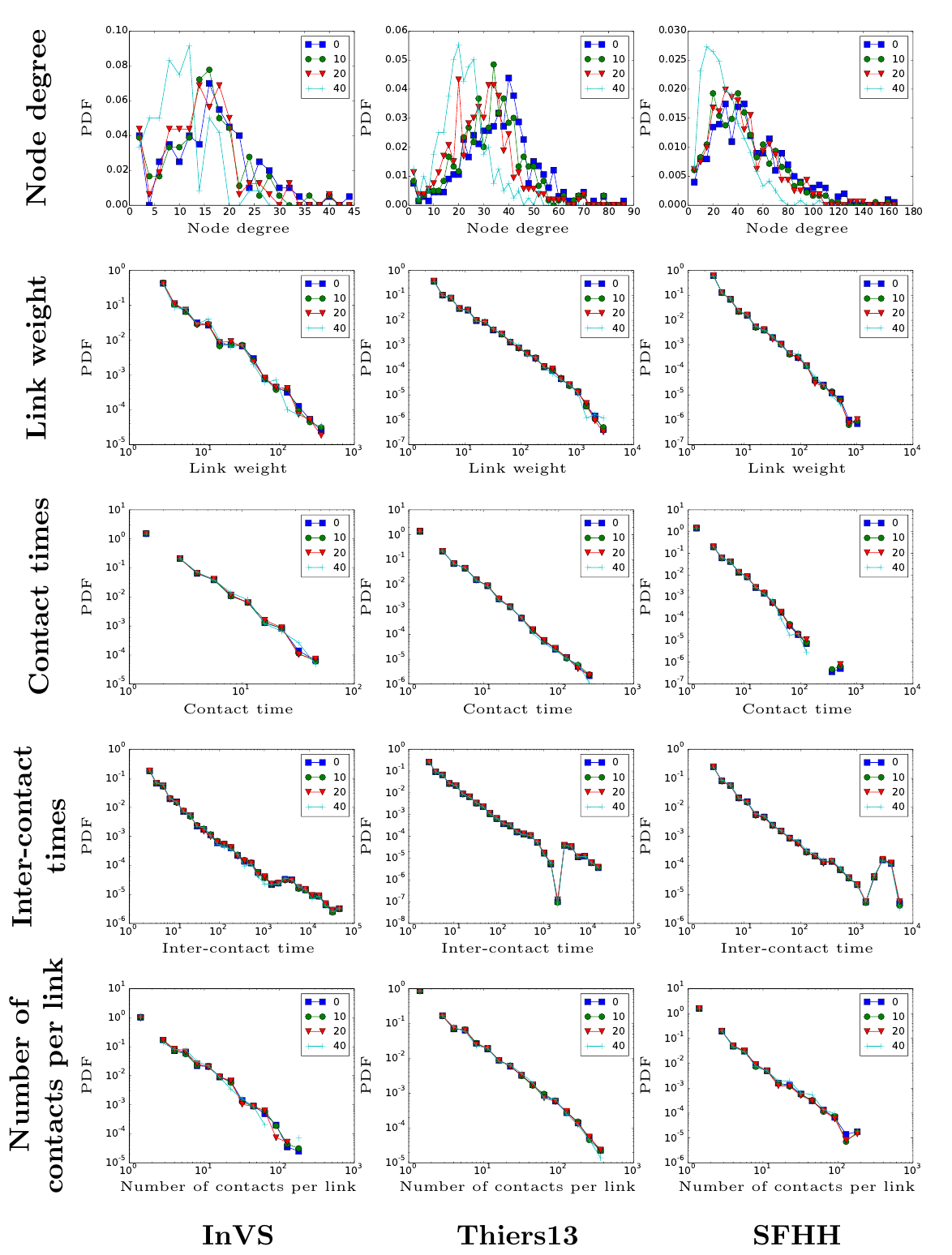}
   \caption{\label{fig:sampling_network}{\bf Effect of sampling on contact network properties.} Comparison of the distributions of
   structural (node degrees and link weights in the aggregated network of contacts) and temporal
   (contact durations, inter-contact times, number of contacts per link) properties of the contact networks,
   for different fractions $f$ of removed nodes.
   For each value of $f$, the distributions are computed on a single realisation of the resampling.}
 \end{figure*}
 
 \begin{figure*}[htp]
   \includegraphics[width=0.95\textwidth]{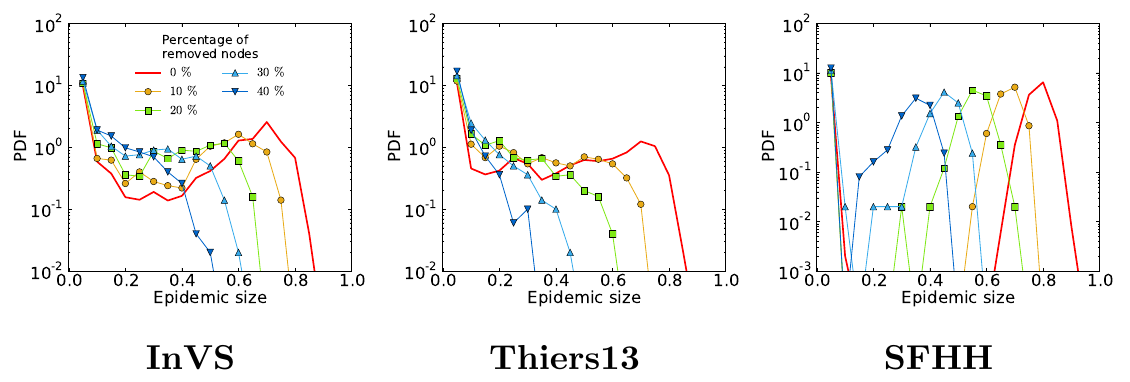}
   \caption{\label{fig:sampling}{\bf Effect of sampling on network density and on the similarity of contact matrices.}
   (Left) Density $\rho$ of the aggregated network of contacts as a function of the fraction $f$ of nodes excluded.
    The shaded areas represent mean $\rho$ $\pm$ s.e.m..
   (Right) Median cosine similarities between the link density contact matrices (CML) of resampled and full data sets, as a function of $f$, for the
   structured populations (high school and workplace). Results are averaged, for each
   value of  $f$, over $1,000$ realisations for the density and over $100$ realisations for the similarities.}
 \end{figure*}

\begin{figure*}[ht]
  \includegraphics[width=0.85\textwidth]{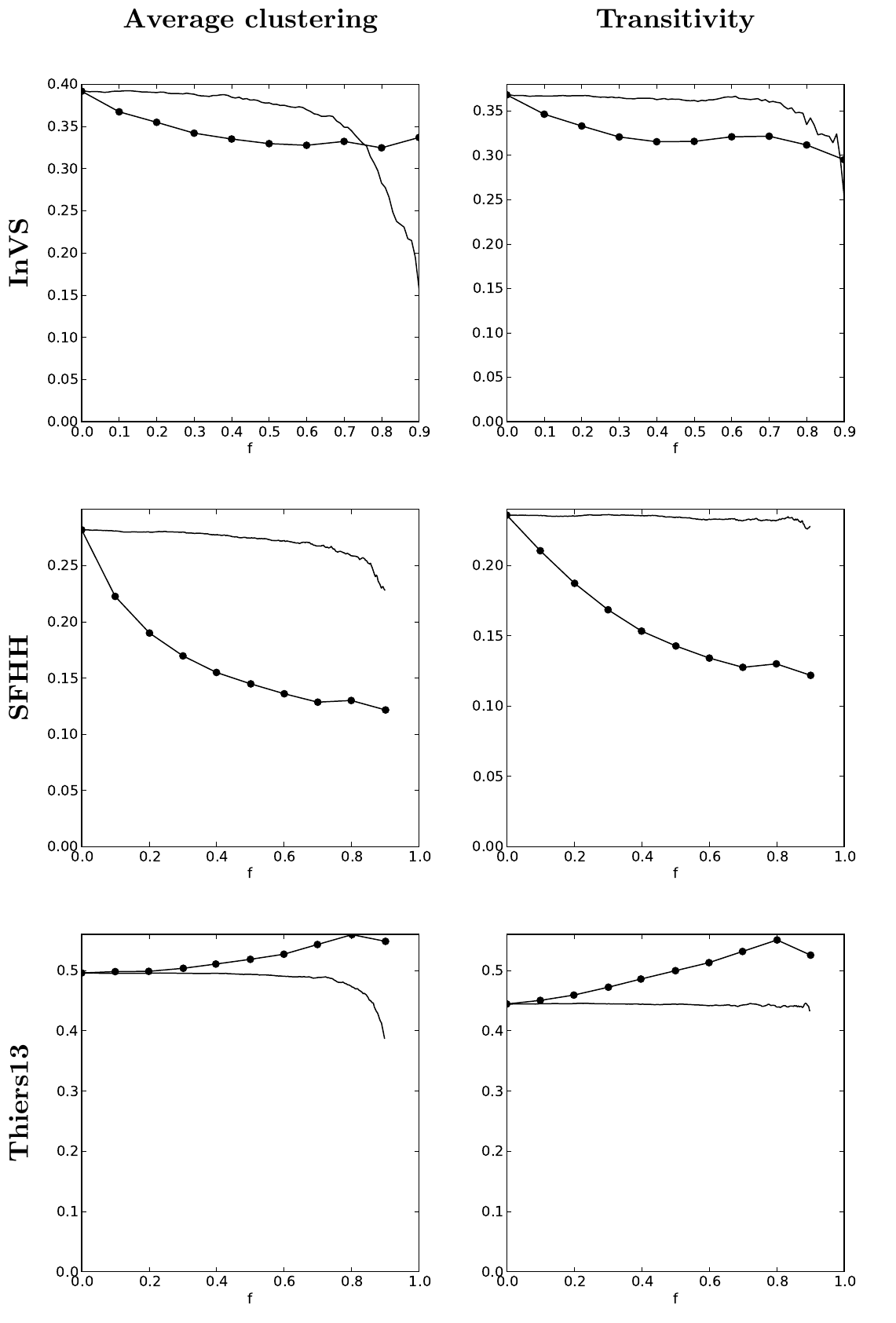}
  \caption{\label{fig:clust_trans}{\bf Effect of sampling and reconstruction
on the average clustering coefficient (left column) and on the network transitivity (right column).} 
The continuous lines show the evolution of clustering coefficient (left) and network transitivity
when the fraction $f$ of removed nodes increases. The full circles correspond to the same quantities
for networks reconstructed using the {\bf WST} method.
Each point is averaged on 100 realisations.}
  \end{figure*}

 \begin{figure*}[htp]
   \includegraphics[width=0.95\textwidth]{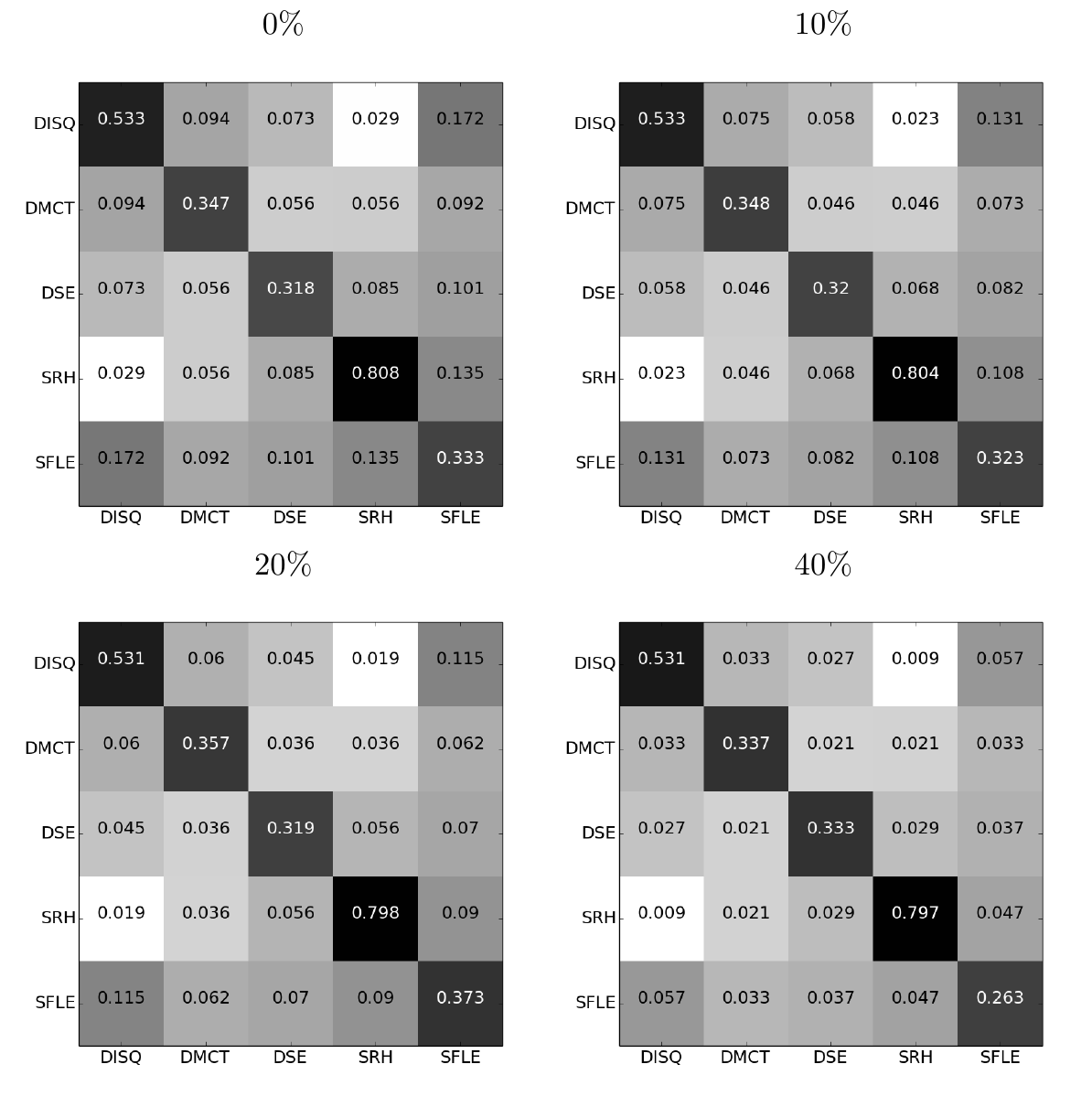}
   \caption{\label{fig:CML_sample_InVS}{\bf Effect of sampling: link density contact matrices ({\em InVS}).}
   Comparison of link density contact matrices for the workplace, for different fractions of excluded nodes, $f$, with the original one ($f=0$).
   Each matrix element $AB$ gives the number of links between nodes of department $A$ and nodes of department $B$ in the contact
   network, normalised by the maximum possible number of such links.
   For each value of $f$, each matrix element is an average over $100$ realisations of the sampling.}
 \end{figure*}

 \begin{figure*}[htp]
   \includegraphics[width=0.95\textwidth]{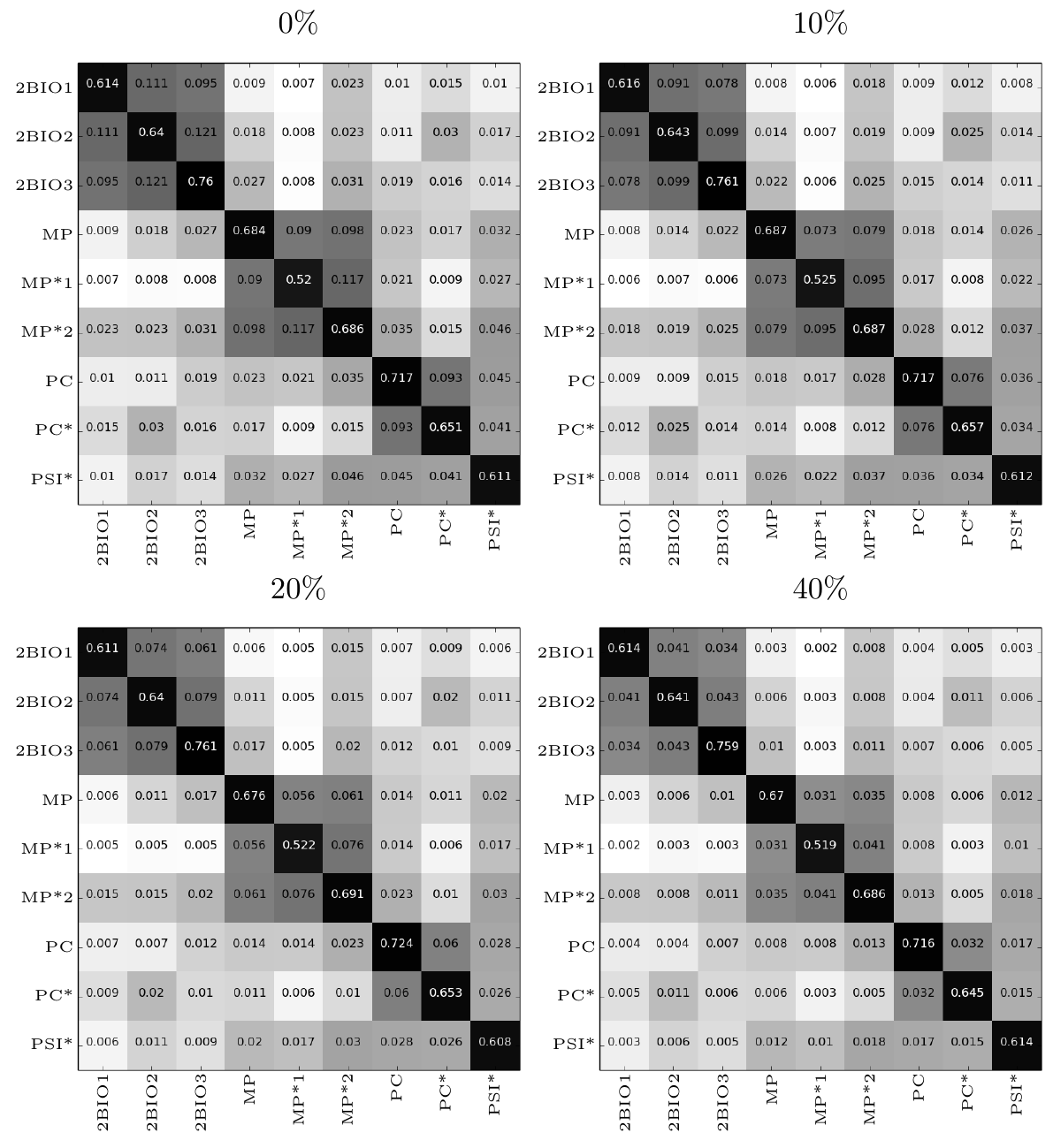}
   \caption{\label{fig:CML_sample_Thiers13}{\bf Effect of sampling: link density contact matrices ({\em Thiers13}).}
   Comparison of the link density contact matrices for the high school, for different fractions $f$ of excluded nodes, with the original one ($f=0$).
   Each matrix element $AB$ gives the number of links between nodes of class $A$ and nodes of class $B$ in the contact
   network, normalised by the maximum possible number of such links.
   For each value of $f$, each matrix element is an average over $100$ realisations of the sampling.}
 \end{figure*}

\begin{figure*}[ht]
  \includegraphics[width=\textwidth]{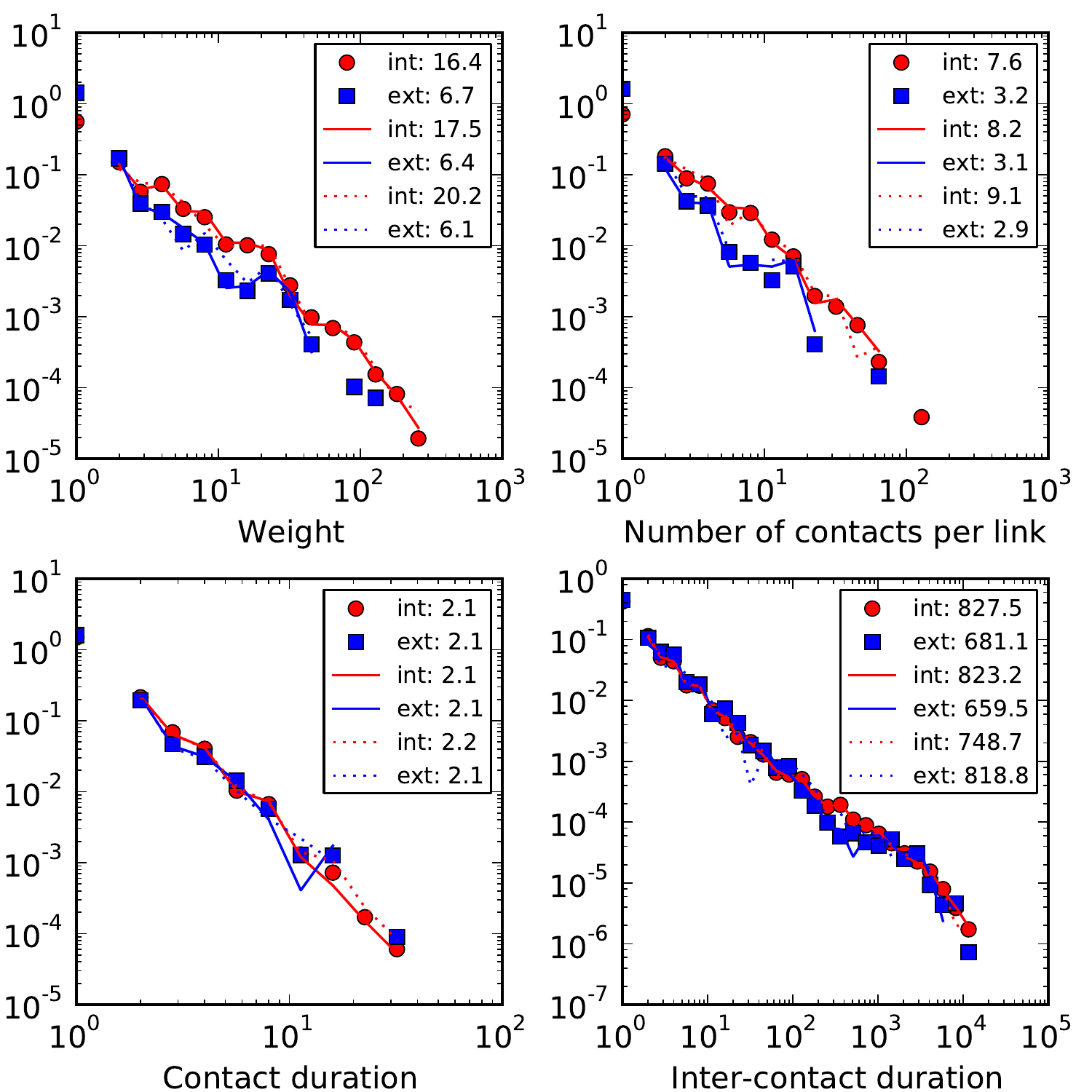}
  \caption{\label{fig:int_ext_InVS}
{\bf Distributions of temporal characteristics for internal (within groups)
and external (between groups) contacts and links ({\em InVS} data).} 
Symbols are for the original data, full lines for resampled data with $f=20\%$,
dotted lines for $f=40\%$. 
Legends give the average values for each distribution.}
  \end{figure*}

\clearpage

\begin{figure*}[ht]
  \includegraphics[width=\textwidth]{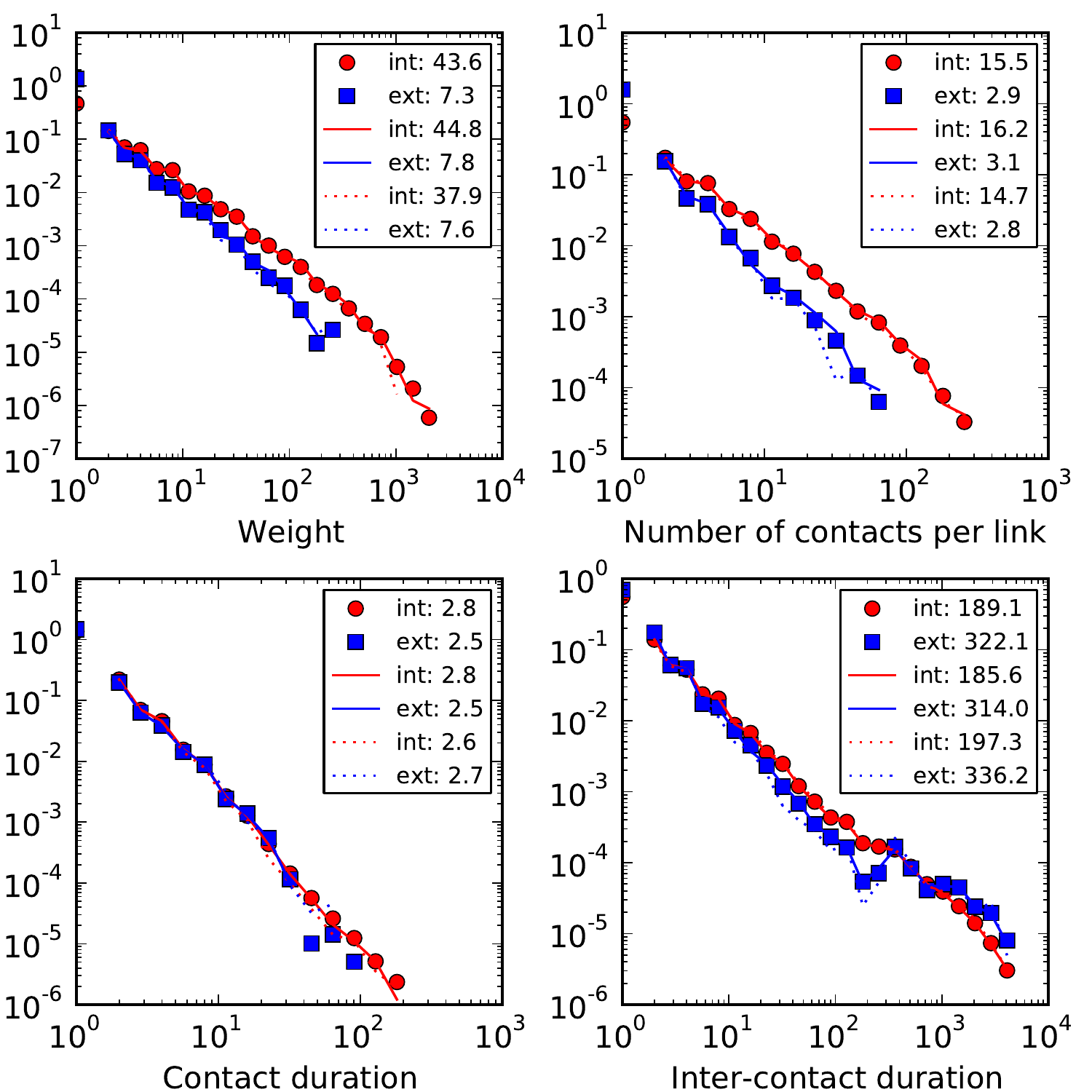}
  \caption{\label{fig:int_ext_Thiers}
{\bf Distributions of temporal characteristics for internal (within groups)
and external (between groups) contacts and links ({\em Thiers13} data).}
Symbols are for the original data, full lines for resampled data with $f=20\%$,
dotted lines for $f=40\%$.
Legends give the average values for each distribution.}
  \end{figure*}

 \begin{figure*}[htp]
   \includegraphics[width=0.95\textwidth]{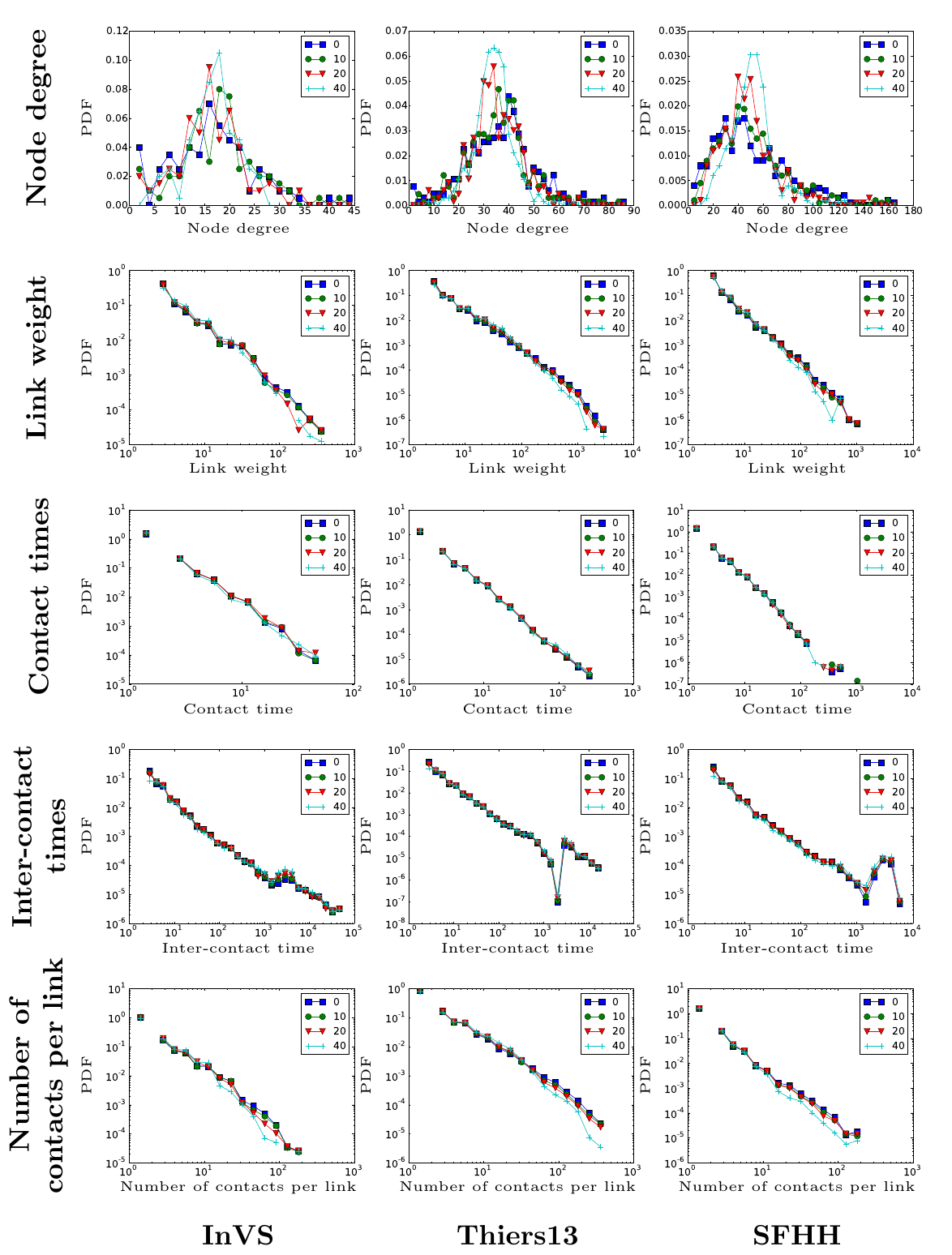}
   \caption{\label{fig:infer_network}{\bf Properties of the reconstructed contact network.}
   Same as Fig. \ref{fig:sampling_network} but for the reconstructed networks:
   Distributions of structural (degrees and weights in the aggregated contact network)
   and temporal (contact times, inter-contact times, number of contacts per link) properties of the surrogate contact networks,
   for different fractions $f$ of nodes excluded. For each value of $f$, the distributions are computed on a single reconstructed network.}
 \end{figure*}
 
  \begin{figure*}[htp]
   \includegraphics[width=0.95\textwidth]{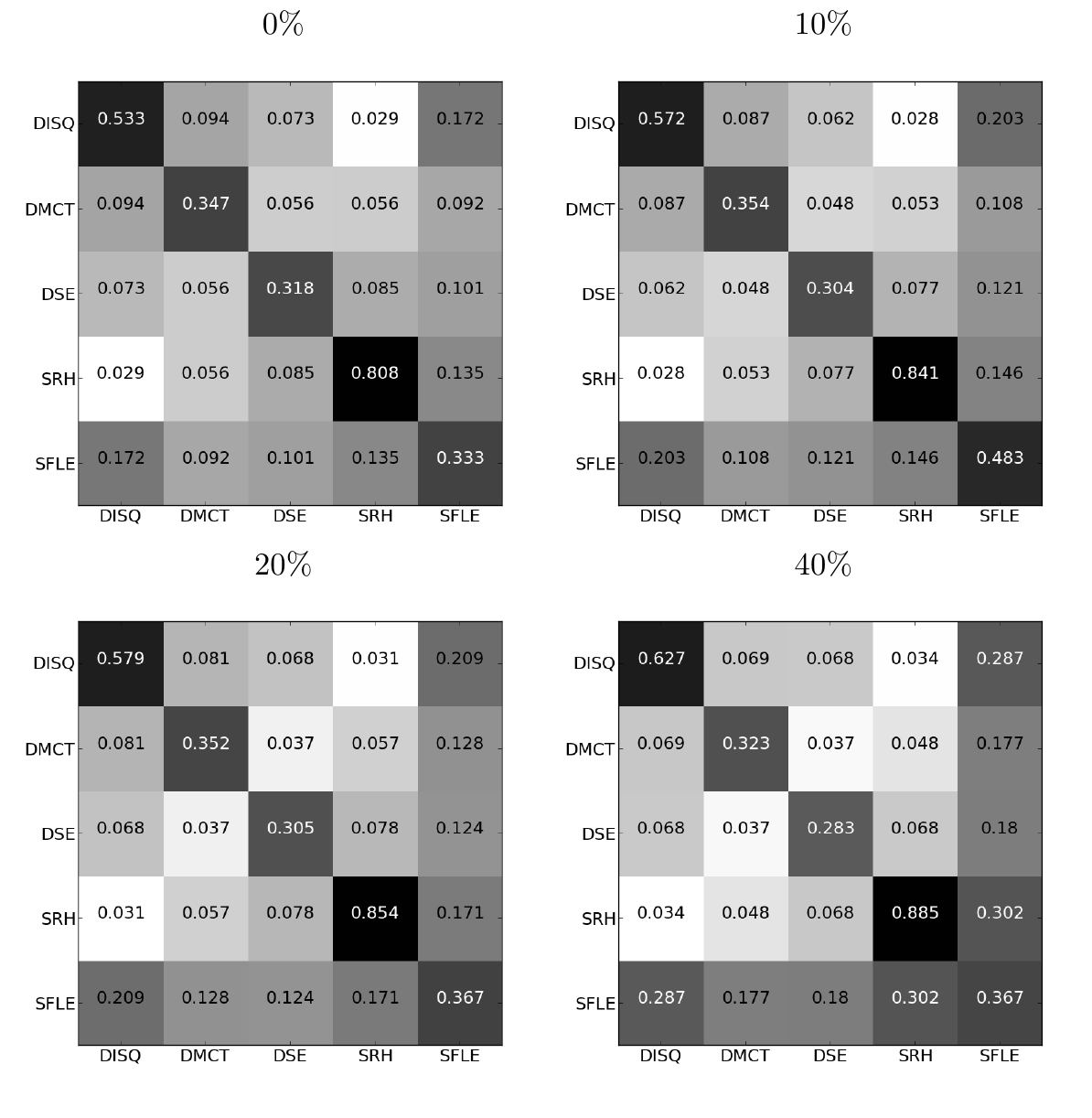}
   \caption{\label{fig:CML_infer_InVS}{\bf Properties of the reconstructed contact network: link density contact matrices ({\em InVS}).}
   Comparison of link density contact matrices for the reconstructed network of the workplace data,
   for different values of the fraction $f$ of excluded nodes, with the original one ($f=0$).
   For each value of $f$, each matrix element is an average over $100$ realisations of the sampling.}
 \end{figure*}

 \begin{figure*}[htp]
   \includegraphics[width=0.95\textwidth]{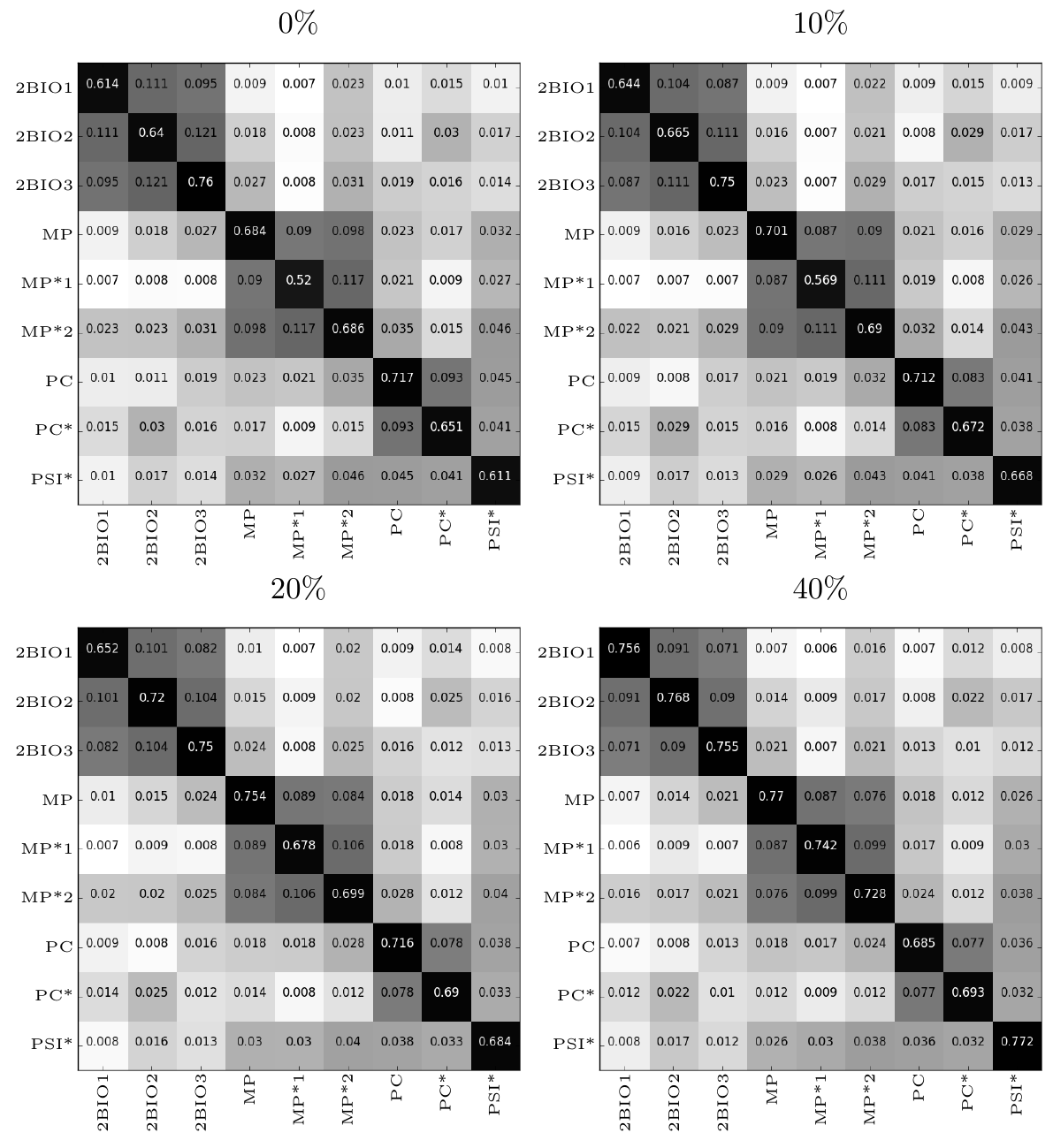}
   \caption{\label{fig:CML_infer_Thiers13}{\bf Properties of the reconstructed contact network: link density contact matrices ({\em Thiers13}).}
   Comparison of link density  contact matrices for the reconstructed network of the high school data, for different values of the fraction $f$ of
   excluded nodes, with the original one ($f=0$).
   For each value of $f$, each matrix element is an average over $100$ realisations of the sampling.}
 \end{figure*}

 \begin{figure*}[htp]
   \includegraphics[width=0.95\textwidth]{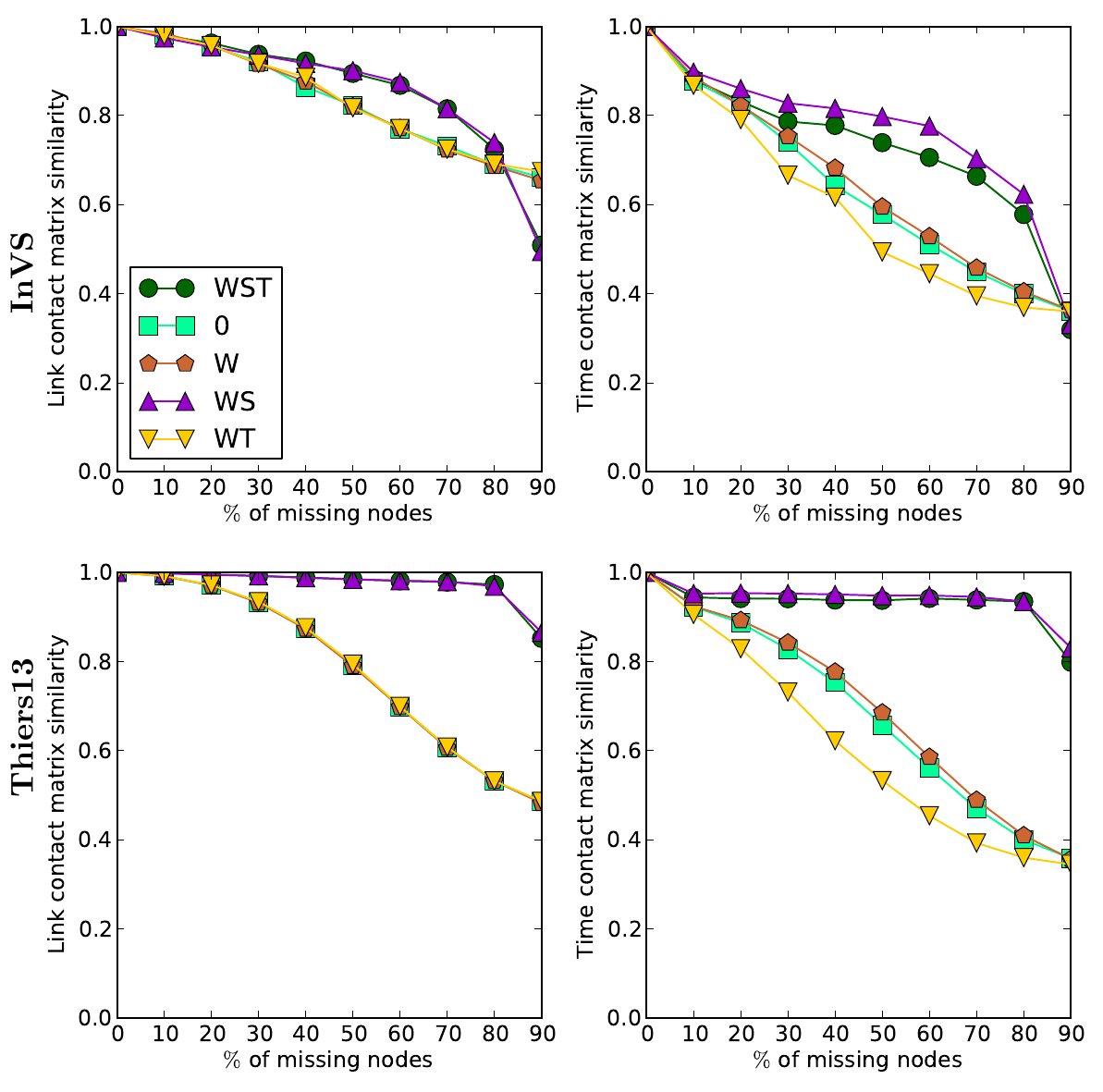}
    \caption{\label{fig:simCM}
    {\bf Similarity of contact matrices for different reconstruction methods}
    Median cosine similarity between the link density and contact time density contact matrices computed between
   the reconstructed network and for the original contact matrices, as a function of the fraction $f$ of removed nodes.
 For each value of  $f$,  the median is computed over $100$ realisations of the reconstruction.
    }
     \end{figure*}

 \begin{figure*}[htp]
   \includegraphics[width=0.95\textwidth]{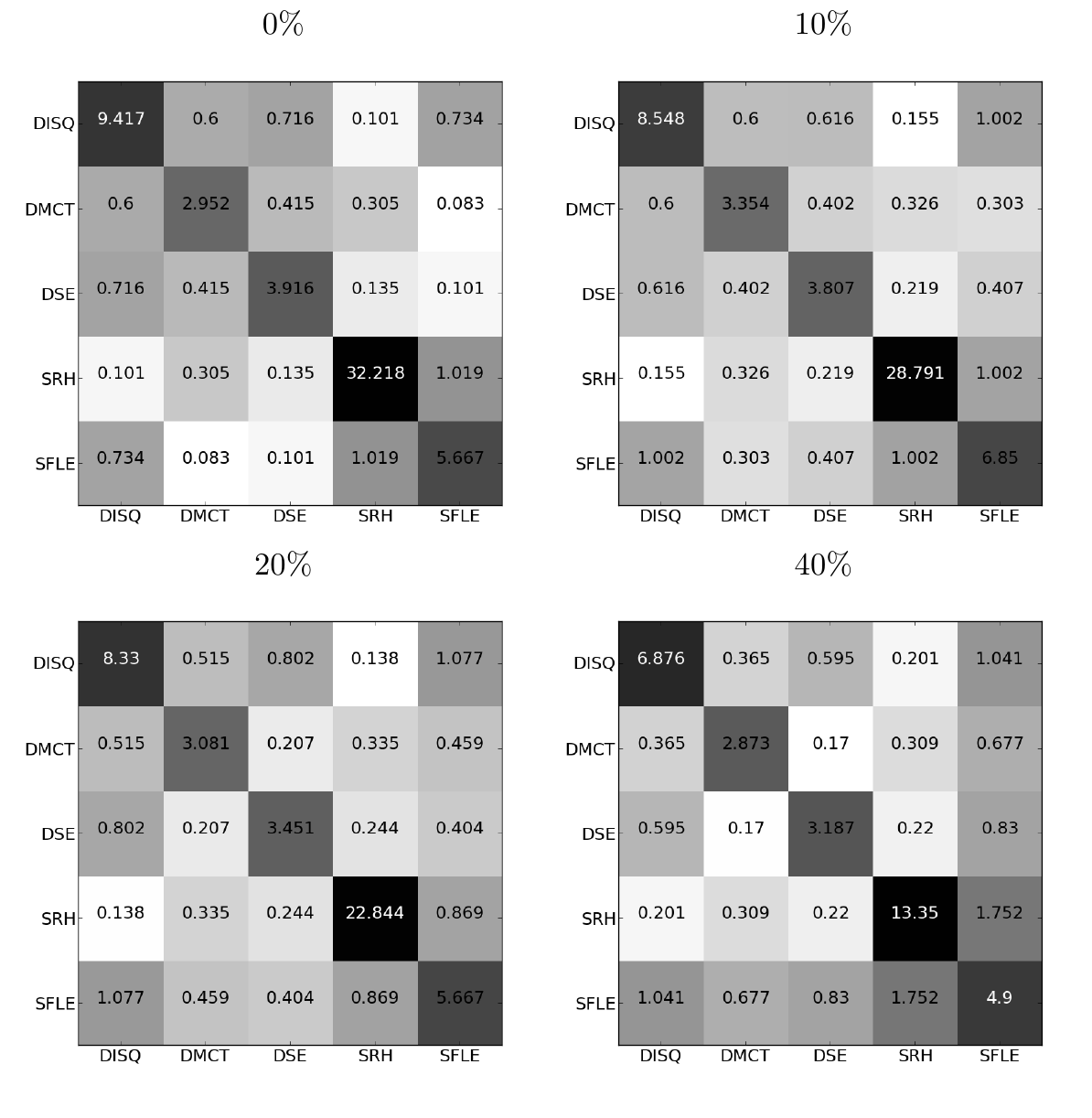}
   \caption{\label{fig:CMT_infer_InVS}{\bf Properties of the reconstructed contact network: time density contact matrices ({\em InVS}).}
   Comparison of the contact time density contact matrices for the reconstructed network of the workplace data, for different fractions of excluded nodes, $f$,
   with the original one ($f=0$).
   Each matrix element $AB$ gives the average time spent in contact between a node of department $A$ and a node of department $B$.
   For each value of $f$, each matrix element is an average over $100$ realisations of the sampling.}
 \end{figure*}

 \begin{figure*}[htp]
   \includegraphics[width=0.95\textwidth]{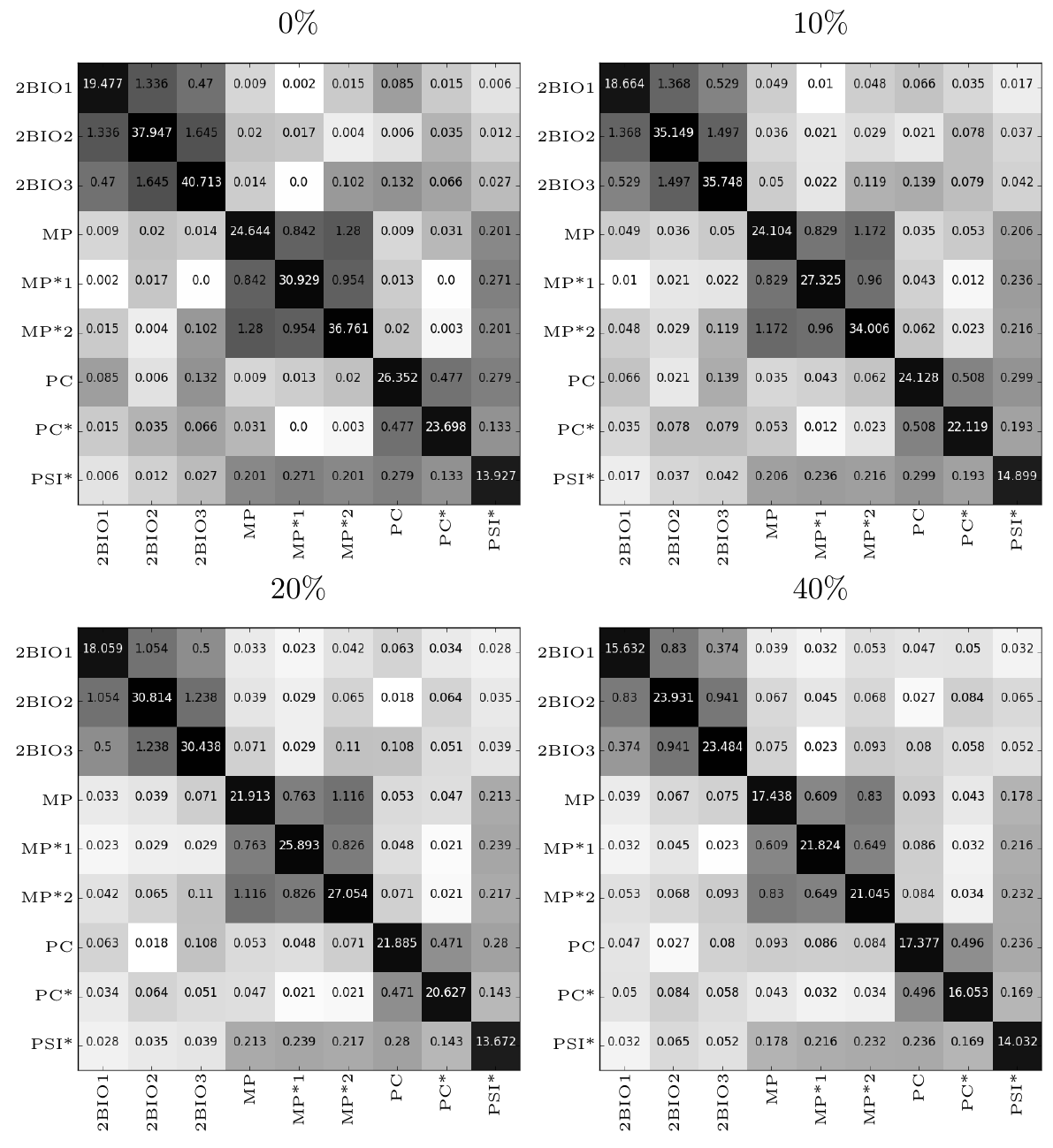}
   \caption{\label{fig:CMT_infer_Thiers13}{\bf Properties of the reconstructed contact network:  time density contact matrices ({\em Thiers13}).}
   Contact time density contact matrices for the reconstructed network of the high school data, for different fractions of nodes excluded, $f$.
     Each matrix element $AB$ gives the average time spent in contact between a node of class $A$ and a node of class $B$.
   For each value of $f$, each matrix element is an average over $100$ realisations of the sampling.}
 \end{figure*}
 
  \begin{figure*}[htp]
   \includegraphics[width=0.95\textwidth]{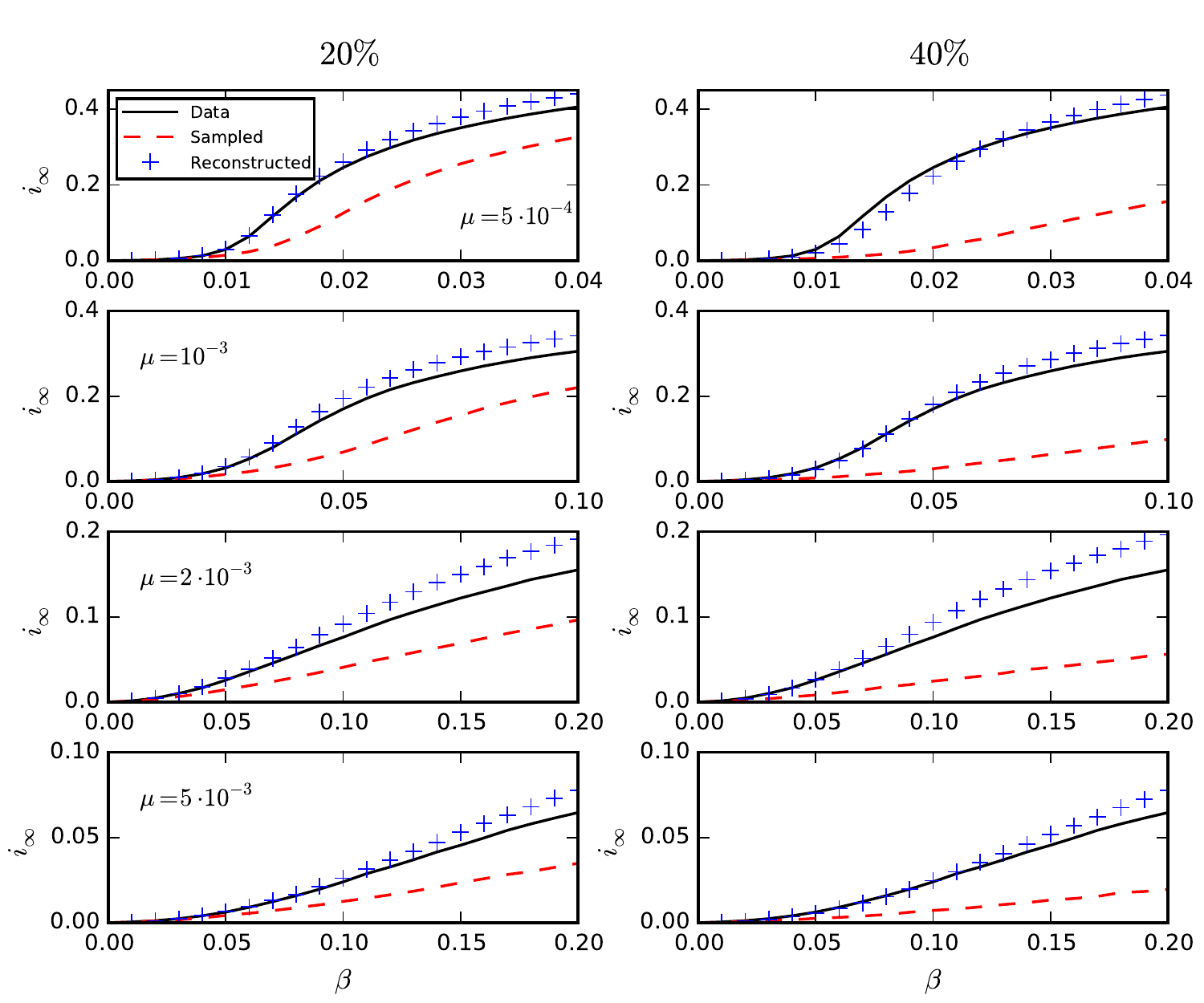}
   \caption{\label{fig:SIS_Thiers}{\bf  Phase diagram of the SIS model for original, resampled and reconstructed
 contact networks ({\em Thiers13} data set). }
 Each panel shows the stationary value $i_\infty$ of the prevalence in the stationary state
 of the SIS model, computed as described in the Methods section, as a function of $\beta$, for several values of $\mu$. Here we consider
 the example of the {\em Thiers13} data set. The epidemic threshold corresponds to the transition between
 $i_\infty=0$ and $i_\infty > 0$. The prevalence curves are computed in each case
 using either the whole data set (continuous lines), resampled data (dashed lines) or reconstructed contact networks (pluses).
 The fraction of excluded nodes in the resampling is $f=20\%$ for the left column and  $f=40\%$ for the right column.
 }
 \end{figure*}

 \begin{figure*}[htp]
   \includegraphics[width=0.95\textwidth]{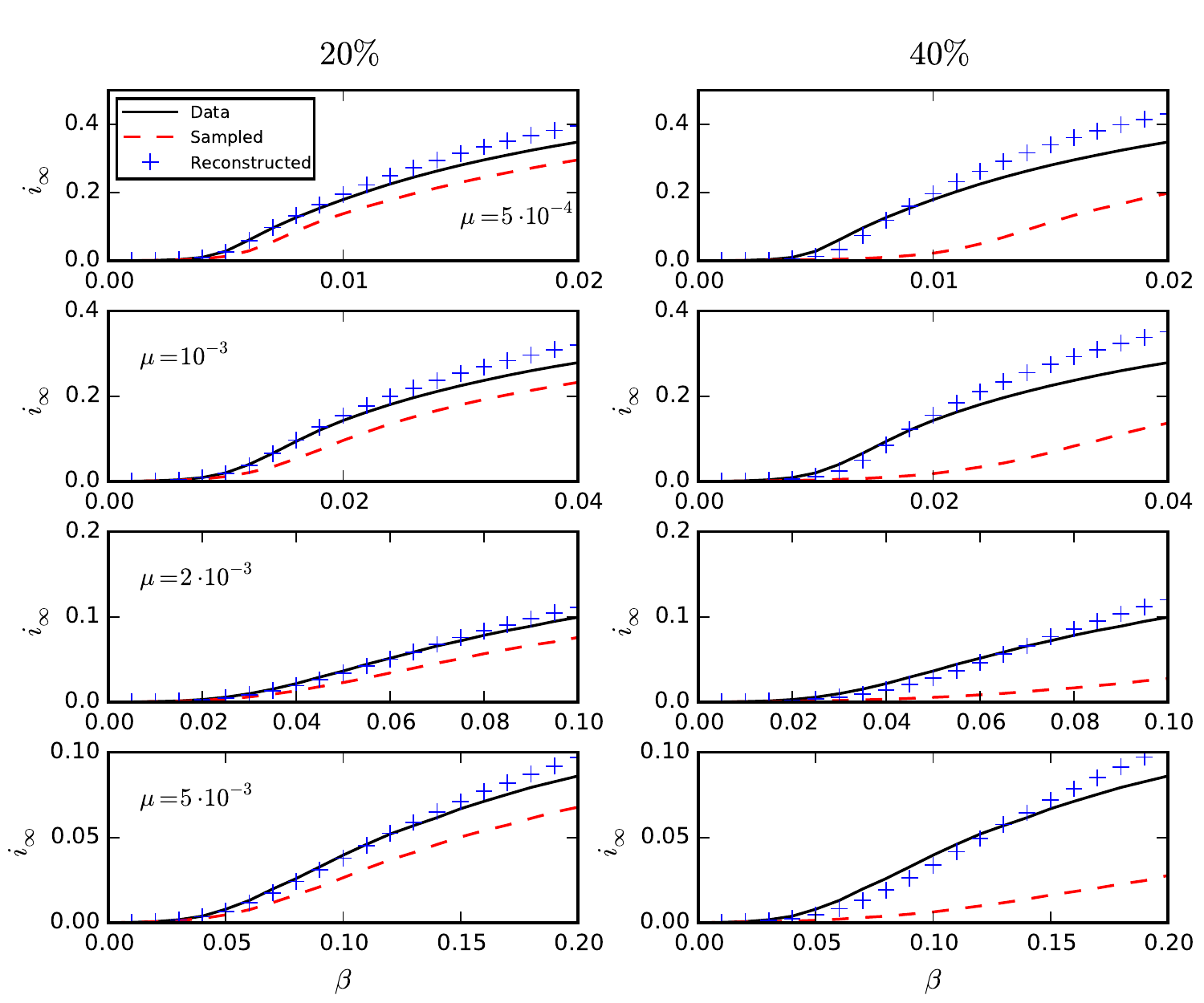}
   \caption{\label{fig:SIS_SFHH}{\bf  Phase diagram of the SIS model for original, resampled and reconstructed
 contact networks ({\em SFHH} data set).}
 Same as Fig.~\ref{fig:SIS_Thiers} for the {\em SFHH} (conference) data set.}
 \end{figure*}

 \begin{figure*}[htp]
  \includegraphics[width=0.95\textwidth]{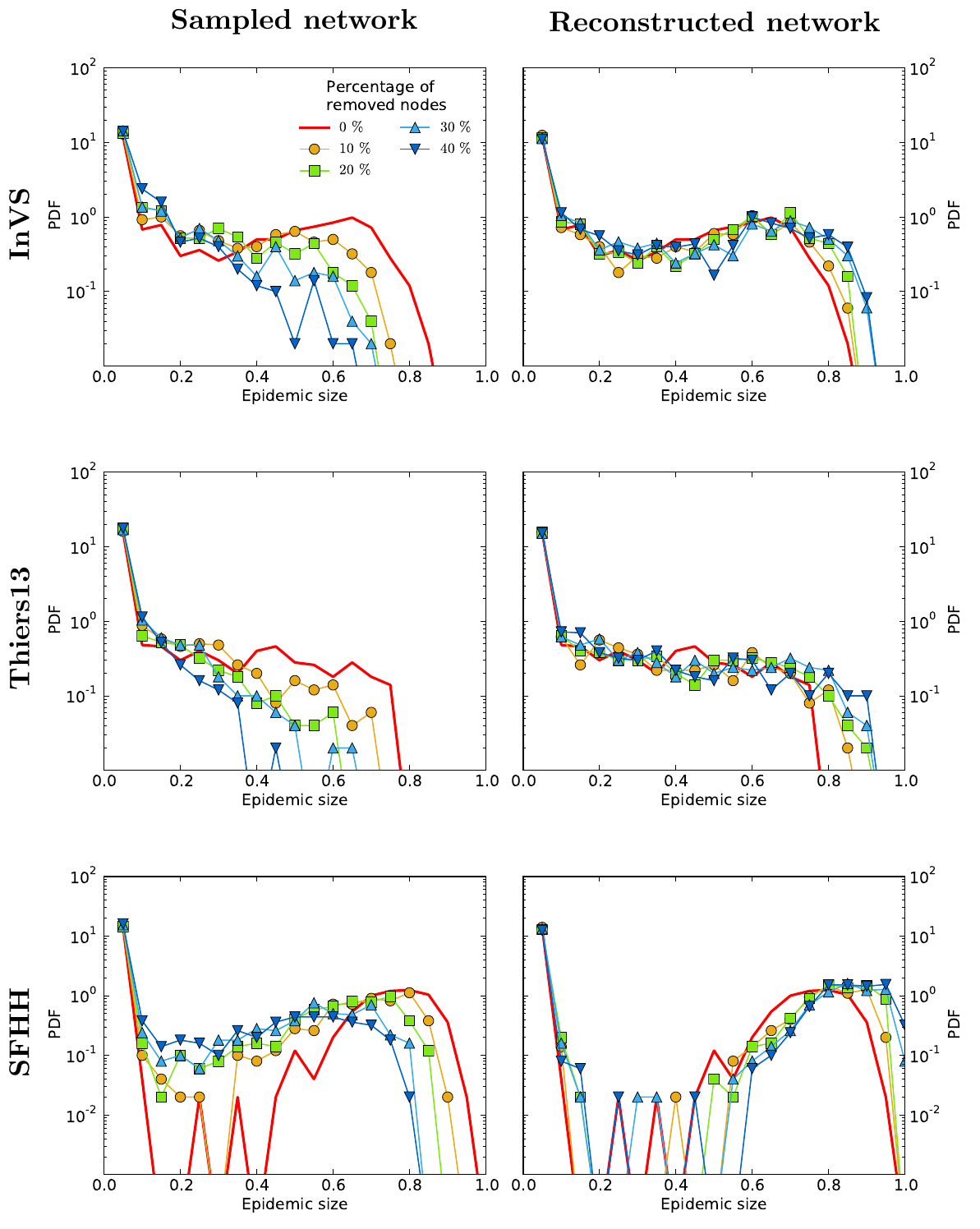}
  \caption{\label{fig:beta.004}{\bf Outcome of SIR epidemic simulations on resampled and
  reconstructed networks for different parameter values.} Distribution of epidemic sizes
(fraction of recovered individuals) at the end of SIR processes simulated on top
  of either resampled (left column) or reconstructed  (right) contact networks, using the {\bf WST} method,
   for different values of the fraction $f$ of nodes removed. The parameters of the
  SIR models are $\beta = 0.004$ and $\beta/\mu = 1000$ ({\em InVS}) or $\beta/\mu = 100$ ({\em Thiers13} and {\em SFHH}).
  The case $f=0$ corresponds to simulations using the whole data set, i.e., the reference case.
  For each value of $f$, $1,000$  independent simulations were performed.}
\end{figure*}
\clearpage

\begin{figure*}[htp]
  \includegraphics[width=0.95\textwidth]{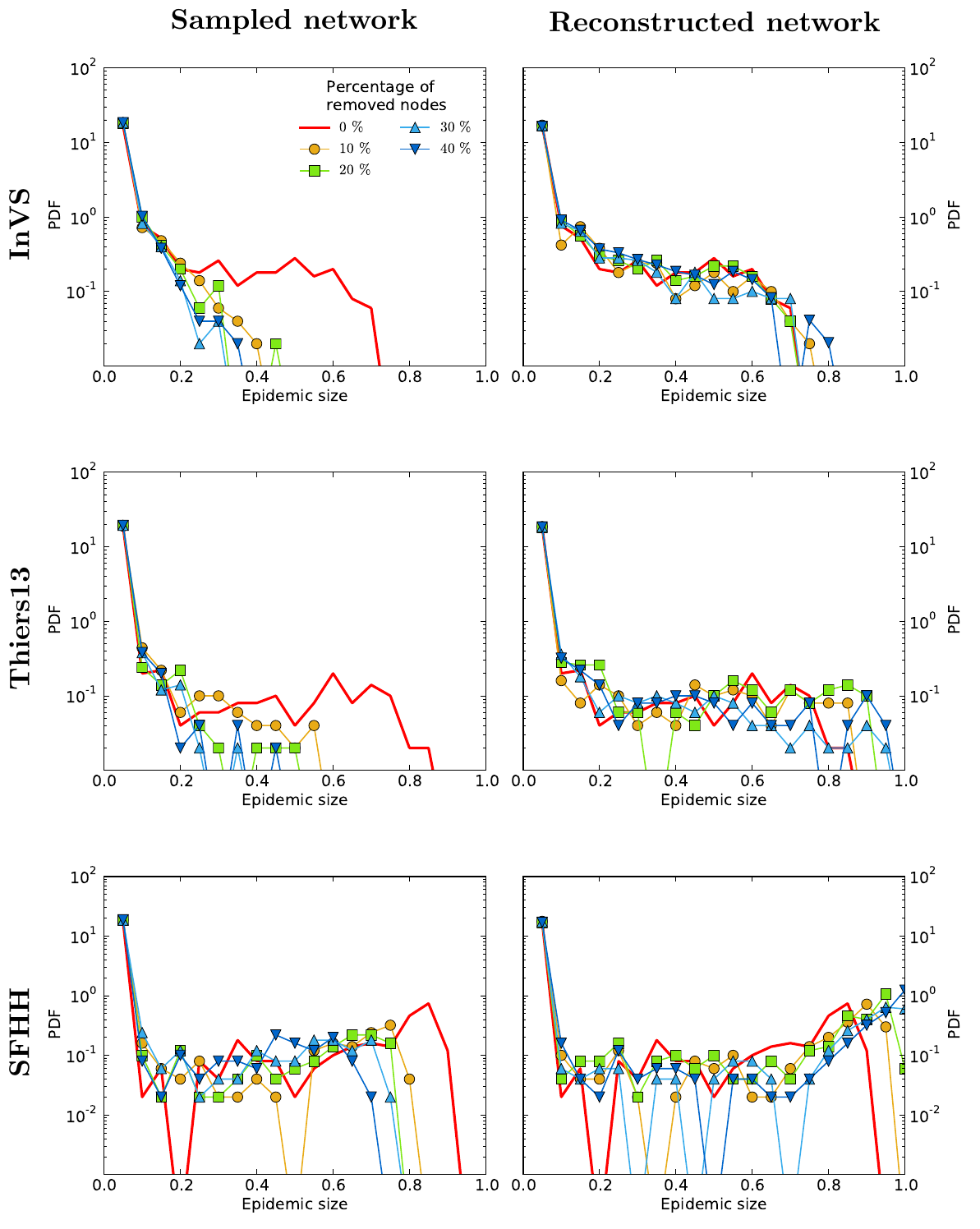}
  \caption{\label{fig:beta.04}{\bf Outcome of SIR epidemic simulations on resampled and
  reconstructed networks for different parameter values.} Distribution of epidemic sizes
(fraction of recovered individuals) at the end of SIR processes simulated on top
  of either resampled (left column) or reconstructed  (right) contact networks, using the {\bf WST} method,
   for different values of the fraction $f$ of nodes removed. The parameters of the
  SIR models are $\beta = 0.04$ and $\beta/\mu = 1000$ ({\em InVS}) or $\beta/\mu = 100$ ({\em Thiers13} and {\em SFHH}).
  The case $f=0$ corresponds to simulations using the whole data set, i.e., the reference case.
  For each value of $f$, $1,000$  independent simulations were performed.}
\end{figure*}

\begin{figure*}[htp]
  \includegraphics[width=0.95\textwidth]{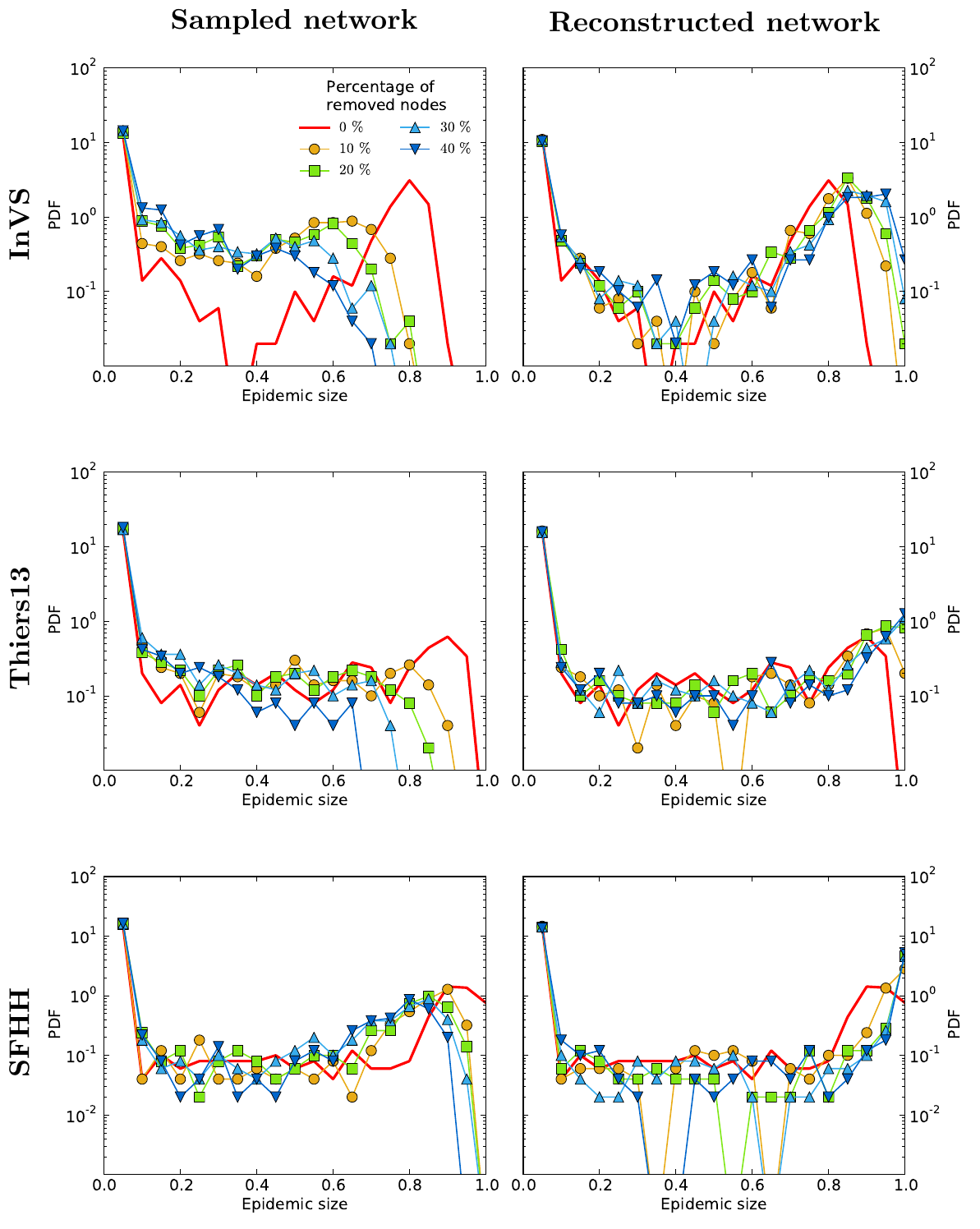}
  \caption{\label{fig:beta.04_large}{\bf Outcome of SIR epidemic simulations on resampled and
  reconstructed networks for different parameter values.  } Distribution of epidemic sizes
(fraction of recovered individuals) at the end of SIR processes simulated on top
  of either resampled (left column) or reconstructed  (right) contact networks, using the {\bf WST} method,
   for different values of the fraction $f$ of nodes removed. The parameters of the
  SIR models are $\beta = 0.04$ and $\beta/\mu = 4000$ ({\em InVS}) or $\beta/\mu = 400$ ({\em Thiers13} and {\em SFHH}).
  The case $f=0$ corresponds to simulations using the whole data set, i.e., the reference case.
  For each value of $f$, $1,000$  independent simulations were performed.}
\end{figure*}

\begin{figure*}[htp]
  \includegraphics[width=0.95\textwidth]{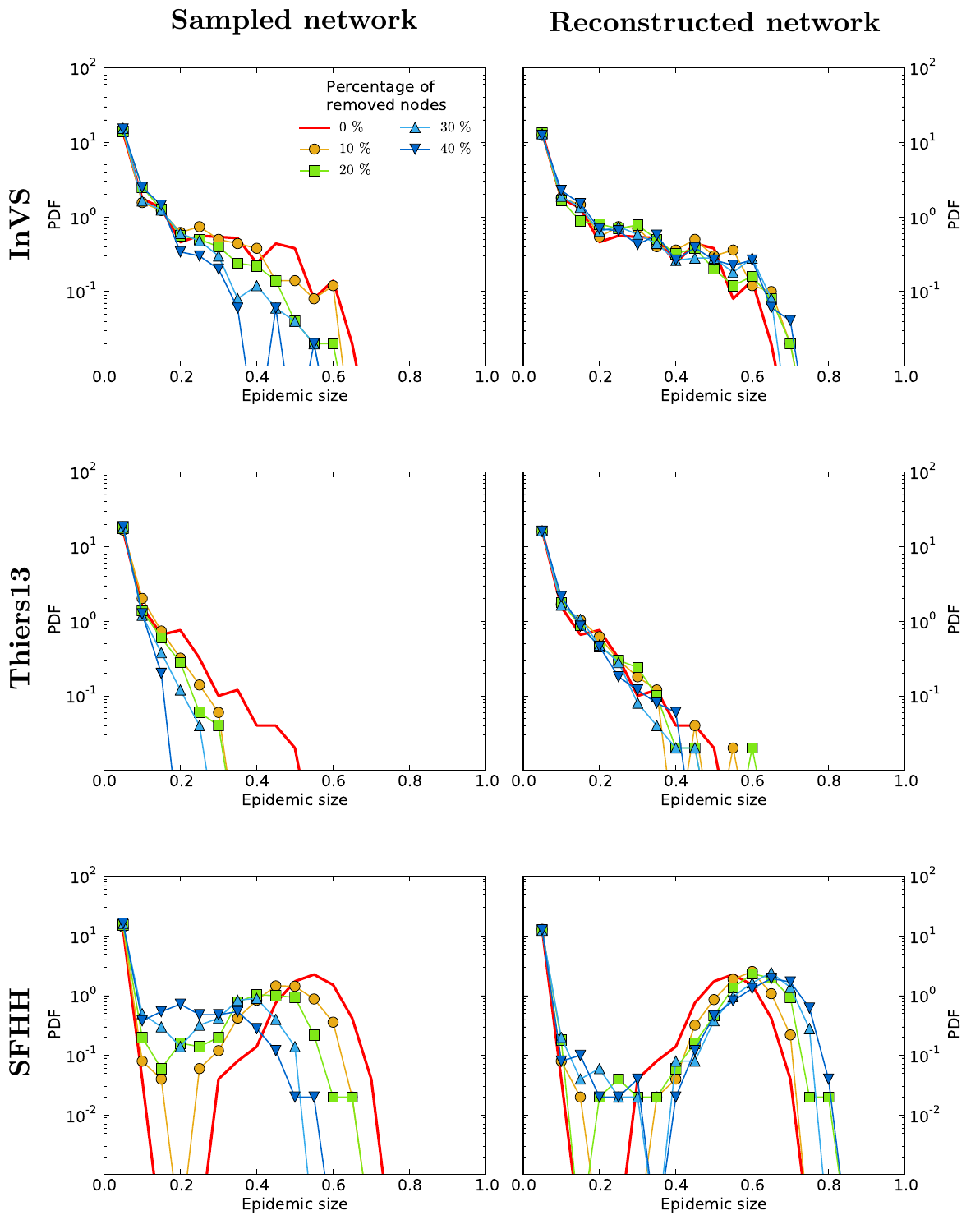}
  \caption{\label{fig:beta.004_small}{\bf Outcome of SIR epidemic simulations on resampled and
  reconstructed networks for different parameter values.  } Distribution of epidemic sizes
(fraction of recovered individuals) at the end of SIR processes simulated on top
  of either resampled (left column) or reconstructed  (right) contact networks, using the {\bf WST} method,
   for different values of the fraction $f$ of nodes removed. The parameters of the
  SIR models are $\beta = 0.0004$ and $\beta/\mu = 500$ ({\em InVS}) or $\beta/\mu = 50$ ({\em Thiers13} and {\em SFHH}).
  The case $f=0$ corresponds to simulations using the whole data set, i.e., the reference case.
  For each value of $f$, $1,000$  independent simulations were performed.}
\end{figure*}

\begin{figure*}[htp]
  \includegraphics[width=0.95\textwidth]{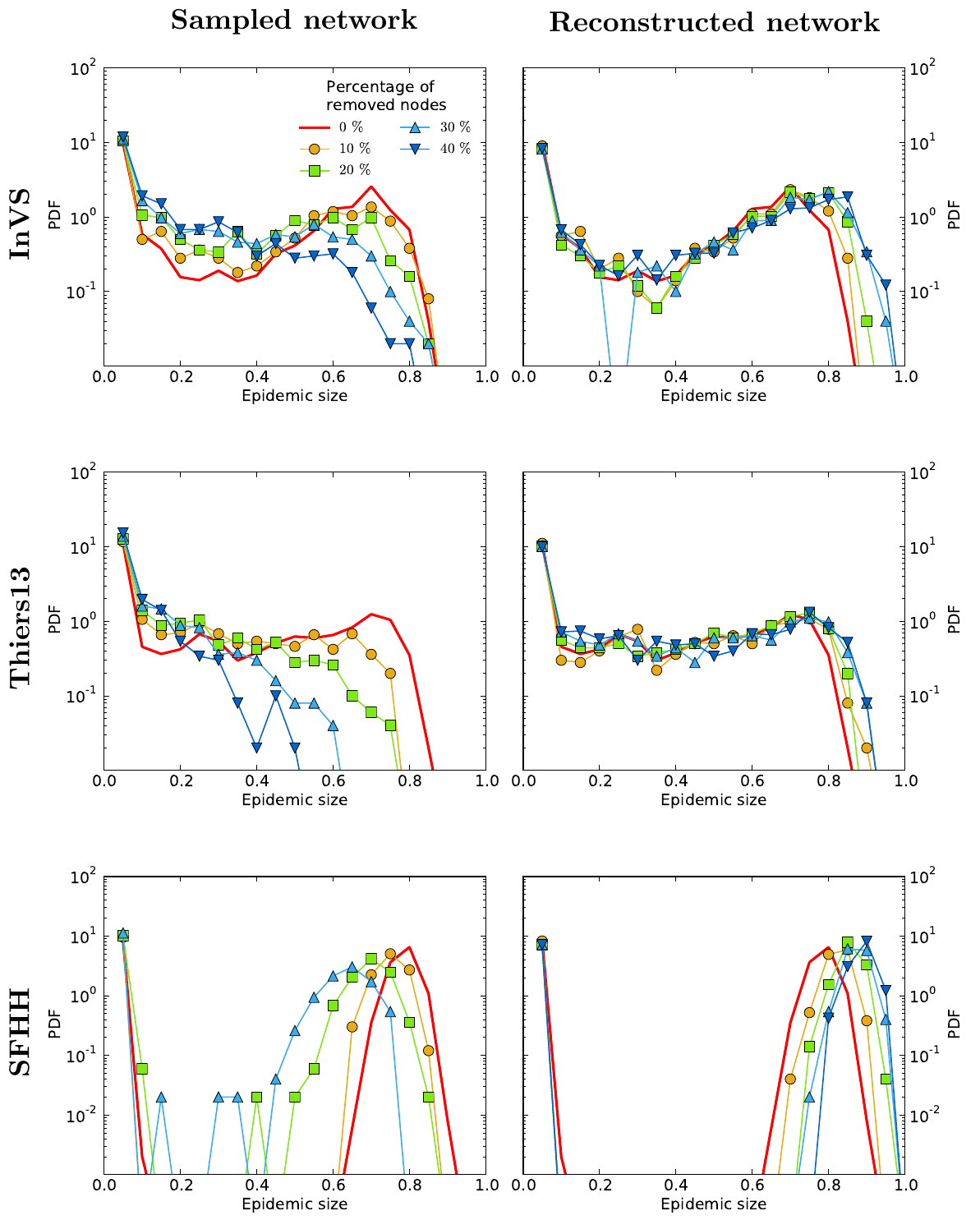}
  \caption{\label{fig:infer_transitivity}{\bf WST method with constrained transitivity. 
Comparison of the outcome of SIR epidemic simulations performed
  on resampled and reconstructed contact networks.} Distribution of epidemic sizes
(fraction of recovered individuals) at the end of SIR processes simulated on top
  of either resampled (left column) or reconstructed  (right) contact networks,
   for different values of the fraction $f$ of nodes removed. The parameters of the
  SIR models are $\beta = 0.0004$ and $\beta/\mu = 1000$
 ({\em InVS}) or $\beta/\mu = 100$ ({\em Thiers13} and {\em SFHH}).
  The case $f=0$ corresponds to simulations using the whole data set, i.e., the reference case.
  For each value of $f$, $1,000$  independent simulations were performed.}
\end{figure*}

\begin{figure*}[htp]
  \includegraphics[width=0.95\textwidth]{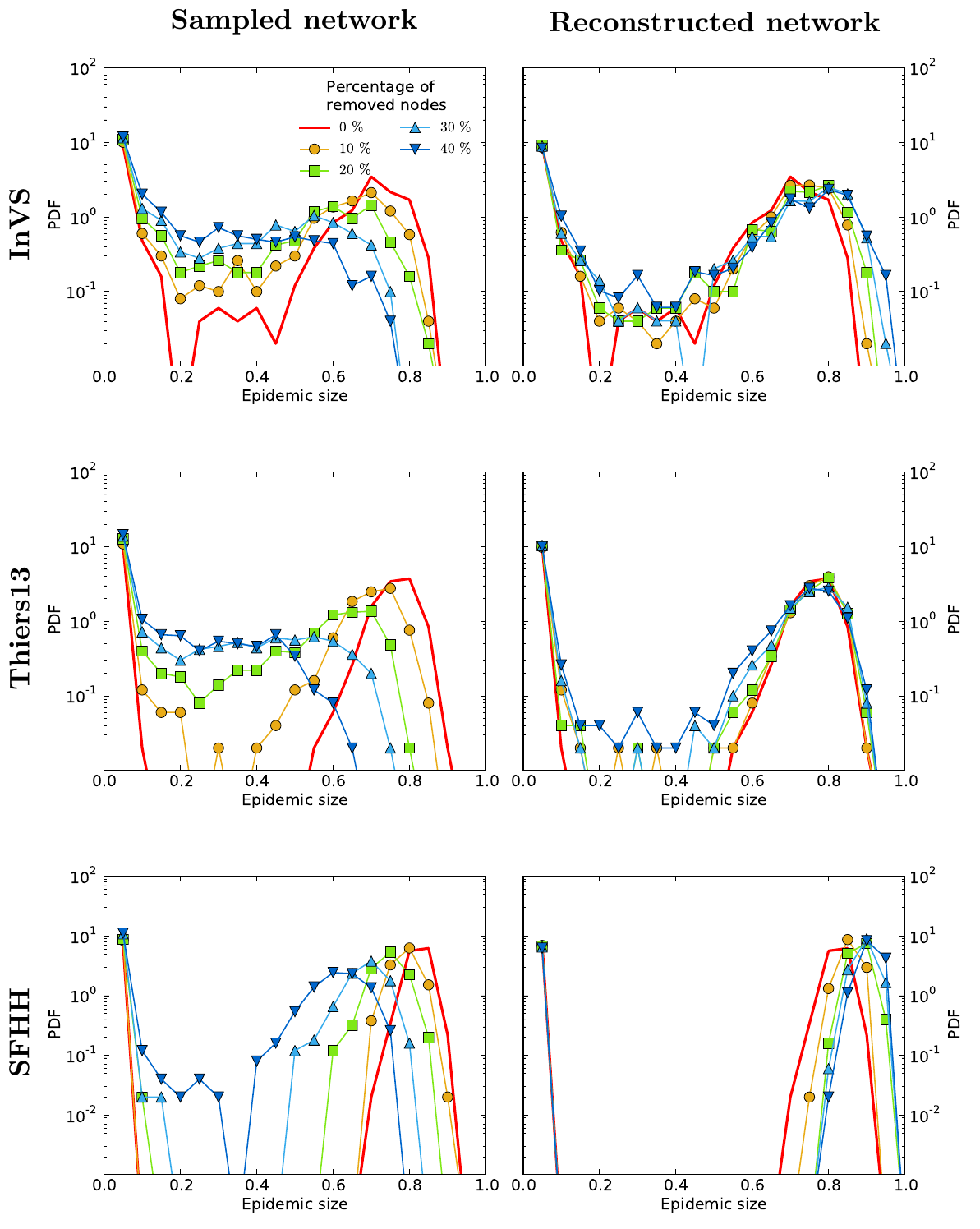}
  \caption{\label{fig:infer_link_shuffle}{\bf Method WST on link-shuffled network. Comparison of the outcome of SIR epidemic simulations performed
  on resampled and reconstructed contact networks.} Distribution of epidemic sizes
(fraction of recovered individuals) at the end of SIR processes simulated on top
  of either resampled (left column) or reconstructed  (right) contact networks,
   for different values of the fraction $f$ of nodes removed. The parameters of the
  SIR models are $\beta = 0.0004$ and $\beta/\mu = 1000$ ({\em InVS}) or $\beta/\mu = 100$ ({\em Thiers13} and {\em SFHH}).
  The case $f=0$ corresponds to simulations using the whole data set, i.e., the reference case.
  For each value of $f$, $1,000$  independent simulations were performed.}
\end{figure*}

\begin{figure*}[htp]
  \includegraphics[width=0.95\textwidth]{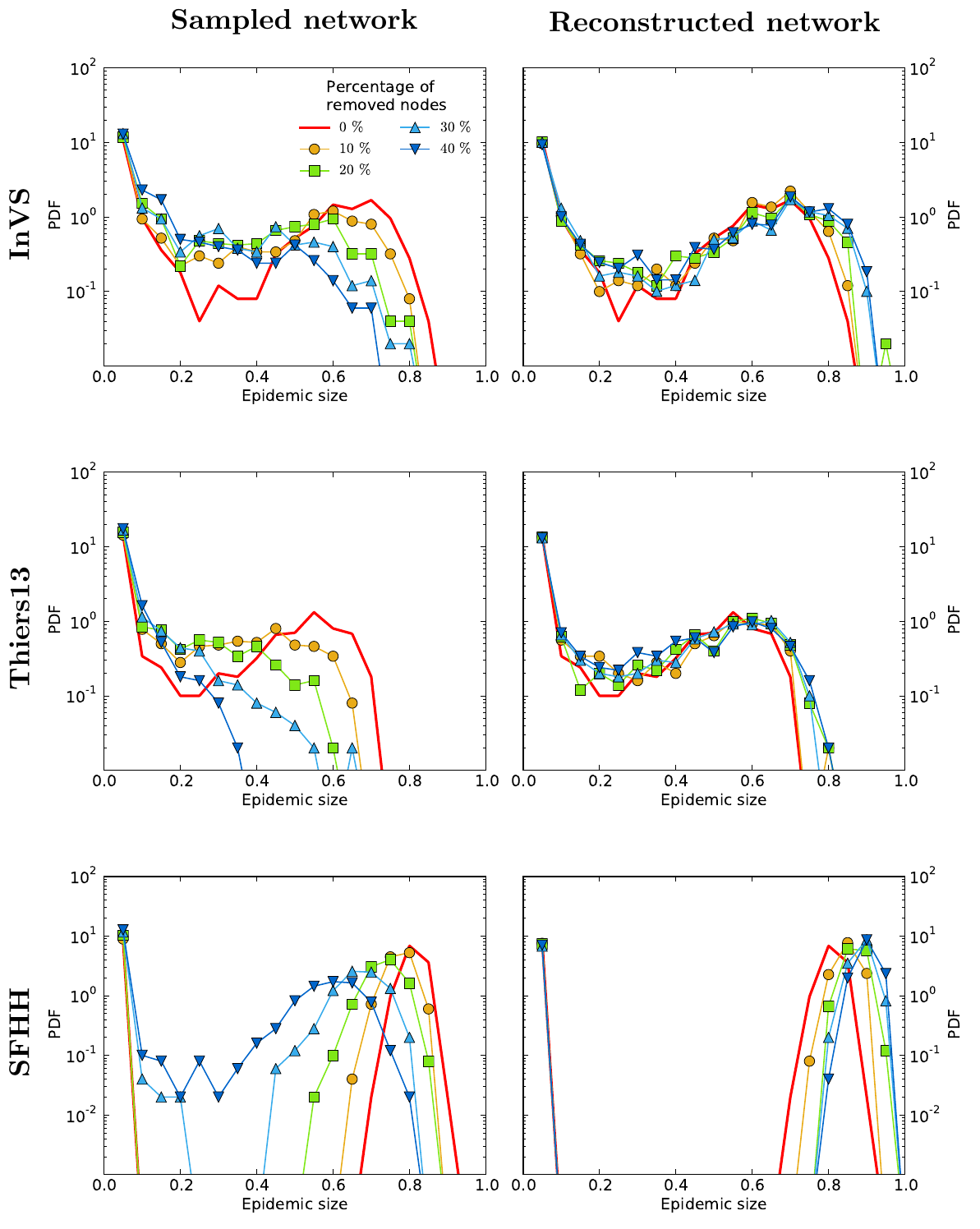}
  \caption{\label{fig:infer_time_shuffle}{\bf Method WST on time-shuffled network. Comparison of the outcome of SIR epidemic simulations performed
  on resampled and reconstructed contact networks.} Distribution of epidemic sizes
(fraction of recovered individuals) at the end of SIR processes simulated on top
  of either resampled (left column) or reconstructed  (right) contact networks,
   for different values of the fraction $f$ of nodes removed. The parameters of the
  SIR models are $\beta = 0.0004$ and $\beta/\mu = 1000$ ({\em InVS}) or $\beta/\mu = 100$ ({\em Thiers13} and {\em SFHH}).
  The case $f=0$ corresponds to simulations using the whole data set, i.e., the reference case.
  For each value of $f$, $1,000$  independent simulations were performed.}
\end{figure*}

\begin{figure*}[htp]
  \includegraphics[width=0.95\textwidth]{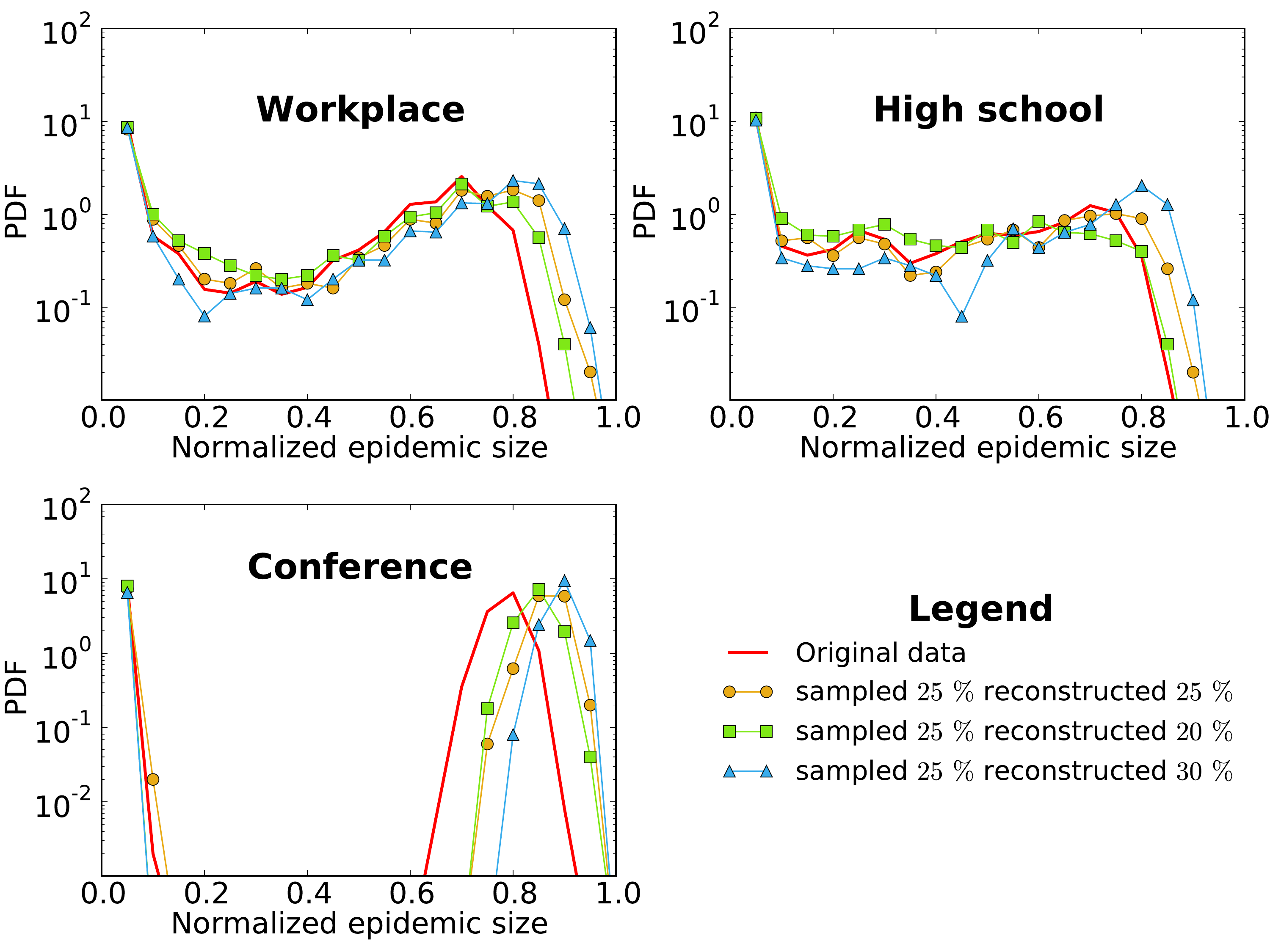}
  \caption{\label{error bar}{\bf Uncertainty on the sampling fraction. Comparison of the outcome of SIR epidemic simulations performed
  on contact networks where $25\,\%$ of nodes were removed, and reconstructed with different values of the assumed
  sampling fraction.} Distribution of epidemic sizes
(fraction of recovered individuals) at the end of SIR processes simulated on top
  of either resampled (left column) or reconstructed  (right) contact networks,
   for different values of the fraction $f$ of nodes removed. The parameters of the
  SIR models are $\beta = 0.0004$ and $\beta/\mu = 1000$ ({\em InVS}) or $\beta/\mu = 100$ ({\em Thiers13} and {\em SFHH}).
  The case ``Original data'' corresponds to simulations using the whole data set, i.e., the reference case.
  For each value of $f$, $1,000$  independent simulations were performed.}
\end{figure*}

\clearpage

\section*{Supplementary notes}

\subsection*{Supplementary note 1: Effect of sampling on the temporal network of contacts}

As described in the main text, we consider temporally resolved networks of contacts ${\cal T}$ in a population ${\cal P}$ of $N$ individuals and
we perform a resampling experiment by selecting a subpopulation ${\cal \tilde{P}}$ of these individuals,
of size $\tilde{N} = (1-f) N$. We assume that only the contacts occurring among the subpopulation ${\cal \tilde{P}}$ are known and
we compare the properties of the corresponding resampled subnetwork ${\cal \tilde{T}}$ with those of the original network.

Supplementary Figure \ref{fig:sampling_network} shows how 
population sampling affects several statistical properties of the contact networks. 
On the one hand, the degree distribution of the aggregated network of contacts
systematically shifts towards smaller degree value. This is expected as each remaining node
has in the resampled network a degree which is at most its degree in the original network, and is strictly smaller if some of its
neighbours are not part of the resampled population. 
On the other hand, the statistical distributions of several quantities of interest are not affected by sampling: This is the case
of the quantities attached either to single contacts or to single links, namely contact and  inter-contact durations, number of contacts per link 
and link weights (the weight of a link is given by the total duration of the contacts between the two corresponding nodes).

Moreover, as shown in Supplementary Figure \ref{fig:sampling}, the density of the aggregated network, {i.e.} the ratio between the number of links and the number of possible links, 
is on average conserved by the random resampling procedure. It varies however 
for different realisations of the resampling, and the corresponding variance increases with the fraction $f$ of excluded nodes.

Supplementary Figure \ref{fig:clust_trans} shows how the average clustering coefficient of the aggregated network
varies with the resampling: notably, it remains high and close to its original value
until large values of $f$ are reached. The transitivity of the network, defined as
three times the number of triangles divided by the number of connected triplets (connected subgraphs of
$3$ nodes and $2$ edges), is even less affected than the clustering coefficient by the resampling procedure.

In the case of structured populations, Supplementary Figures \ref{fig:CML_sample_InVS} \& \ref{fig:CML_sample_Thiers13} show 
that the stability of the resampled network's density holds at the more detailed level of the 
contact matrices of link densities. In such matrices,  the element $(i,j)$ 
is given by the number of links between individuals of groups $i$ and $j$, normalised by the total number of possible links
between these two groups (if $n_i$ denotes the number of individuals in group $i$, the number of possible links is equal to $n_i n_j/2$ for $i \ne j$ and to $n_i(n_i-1)/2$ for $i=j$). These figures clearly illustrate how the diagonal and block-diagonal structures are preserved, and 
Supplementary Figure \ref{fig:sampling} gives a quantitative assessment of this stability by showing that the 
cosine similarity between contact matrices between the resampled and original aggregated contact networks
remains high even for when a large fraction of the nodes are excluded.

We moreover illustrate in Supplementary Figures \ref{fig:int_ext_InVS} and \ref{fig:int_ext_Thiers} the difference in statistical
properties of contacts and links within and between groups, still for structured populations:
\begin{itemize}
\item the distributions of contact durations are indistinguishable;
\item the distribution of link weights (aggregated contact durations) is broader
for links between individuals belonging to the same group than for links joining
individuals of different groups;
\item this is due to the difference in the distributions of numbers of contacts per link, which is
broader for links within groups than for links between groups;
\item the distributions of inter-contact durations differ also slightly, with smaller averages
for within-group links.
\end{itemize}
Most importantly, all these properties and distributions remain stable under resampling, showing that reliable
information on the distributions of contact and inter-contact durations, aggregated contact durations, numbers
of contacts per link, can be obtained in the resampled data, including the statistical
differences between links joining members of different groups and links between two individuals of the same group.

\subsection*{Supplementary note 2: Properties of the reconstructed contact networks}

As described in the main text and in particular in the Methods section, we construct a surrogate set of contacts concerning the $f N$ individuals
excluded by the resampling. We compare here the properties of the resulting contact networks (obtained by merging the resampled contact network ${\cal \tilde{T}}$ 
and the surrogate set of contacts) and of the original contact network, ${\cal T}$.

Supplementary Figure \ref{fig:infer_network} shows that the degree distribution, which is not constrained by the reconstruction procedure, deviates 
from the original distribution. On the other hand, the distributions of contact durations, inter-contact durations, number of contacts per link and
link weights are preserved. 
Moreover, the link density contact matrices of the reconstructed networks (Supplementary Figure \ref{fig:CML_infer_InVS} \& \ref{fig:CML_infer_Thiers13}) share a high similarity with
 the original contact matrices, even for high fractions of nodes excluded (Supplementary Figure \ref{fig:simCM}). 
 
 For completeness, we also compute the contact matrices in contact time density (CMT), in which 
each element $(i,j)$ 
is given by the total time in contact between individuals of groups $i$ and $j$, normalised by the total number of possible links
between these two groups: it gives the average time spent in contact by two random individuals of groups $i$ and $j$.
Supplementary Figures \ref{fig:simCM}, \ref{fig:CMT_infer_InVS} and  \ref{fig:CMT_infer_Thiers13} 
show that the structure of these
matrices is well recovered by the reconstruction methods, with high similarity with the original matrices.

\subsection*{Supplementary note 3: Phase diagram of the SIS model for the conference and  high school data sets}

We observe for the high school and the conference the same effect on the phase diagram of the SIS model as in the workplace: 
sampling leads to a shift of the epidemic threshold to higher values and thus to an underestimation of the epidemic risk. 
The phase diagram and the epidemic threshold are estimated more accurately by using reconstructed networks, 
thus giving a better evaluation of the epidemic risk (Supplementary Figures \ref{fig:SIS_Thiers} \& \ref{fig:SIS_SFHH}).

\subsection*{Supplementary note 4: Sensitivity analysis}

In the main text, we have considered values of the SIR model parameters leading to a non-negligible epidemic risk and a value of $\beta$ corresponding
to slow processes. We consider here several other values of the parameters, corresponding either
to faster processes (Supplementary Figures \ref{fig:beta.004} - \ref{fig:beta.04_large})
or to smaller epidemic risk (Supplementary Figure \ref{fig:beta.004_small}). In all cases, simulations performed on the resampled
contact networks lead to a strong underestimation of the epidemic sizes, with distributions shifting to smaller values as $f$ increases, while the use
of reconstructed data sets leads to a better estimation and generally speaking a slight overestimation of the epidemic risk.

\section*{Supplementary methods}

\subsection*{Detailed alternative reconstruction methods}

We give here details on the alternative reconstruction methods mentioned in the main text, which use less information than the
{\bf WST} method. 
In each case we consider the same setup as the complete method: a population ${\cal P}$ of $N$ individuals 
(the nodes of the contact network), potentially organised in groups, for which we know all the
contacts taking place among a subpopulation ${\cal \tilde{P}}$ of size $\tilde{N} = (1-f) N$.
For  the remaining $n = N - \tilde{N} = f N$ individuals, no contact information is available, but 
we know  to which group they belong. 
We also have access to the overall activity timeline, \emph{i.e.}, to 
the successive intervals during which contacts can happen (daytimes), and are excluded (nights and weekends).
The alternative reconstruction methods are the following:

\begin{description}
\item[0] We perform the reconstruction using only
the network density and the average link weight, both measured in the resampled network ${\cal \tilde{T}}$. The algorithm goes as follows:
  \begin{enumerate}
  \item we measure in the resampled data:
    \begin{itemize}
    \item the density $\rho$ of links in the time-aggregated network;
    \item the average link weight $\langle w \rangle_{\rm s}$ (the weight of a link is defined as the total contact time between the two linked nodes);
    \end{itemize}

  \item we compute the number of links $e$ that must be added to keep the network density constant when we add the $n$ excluded nodes;

  \item we construct $e$ links according to the following procedure:
    \begin{itemize}
    \item a node $i$ is randomly chosen from the set ${\cal P} \backslash {\cal \tilde{P}}$ of excluded nodes;
    \item a node $j$ is randomly chosen from the set ${\cal P} \backslash \{i\}$ of all other nodes;
    \item we compute $n_{ij} = \langle w \rangle_{\rm s} /\Delta t$, where $\Delta t=20s$ is the temporal resolution of the data set, and we
    randomly choose $n_{ij}$ time windows of length $\Delta t$ within the activity windows defined by the activity timeline as contact
    events between $i$ and $j$.
    \end{itemize}
  \end{enumerate}
  
\item[W] We perform the reconstruction using only
the network density and the distribution of link weights, both measured in the resampled network ${\cal \tilde{T}}$. The algorithm goes as follows:
  \begin{enumerate}
  \item we measure in the resampled data:
    \begin{itemize}
    \item the density $\rho$ of links in the time-aggregated network;
    \item the list $\{w\}$ of link weights (the weight of a link is defined as the total contact time between the two linked nodes);
    \end{itemize}

  \item we compute the number of links $e$ that must be added to keep the network density constant when we add the $n$ excluded nodes;

  \item we construct $e$ links according to the following procedure:
    \begin{itemize}
    \item a node $i$ is randomly chosen from the set ${\cal P} \backslash {\cal \tilde{P}}$ of excluded nodes;
    \item a node $j$ is randomly chosen from the set ${\cal P} \backslash \{i\}$ of all other nodes;
    \item from $\{w\}$, we draw the weight $w_{ij}$ of the link $ij$;
    \item we compute $n_{ij} = w_{ij} /\Delta t$, where $\Delta t=20s$ is the temporal resolution of the data set, and we
    randomly choose $n_{ij}$ time windows of length $\Delta t$ within the activity windows defined by the activity timeline as contact
    events between $i$ and $j$.
    \end{itemize}
  \end{enumerate}
  
\item[WS] We perform the reconstruction using the network density, the distributions of link weights for internal 
(within groups) and external (between groups) links, and the structure of the aggregated 
network given by the link density contact matrix, all measured in the resampled network ${\cal \tilde{T}}$. The algorithm goes as follows:
  \begin{enumerate}
  \item we measure in the resampled data:
    \begin{itemize}
    \item the density $\rho$ of links in the time-aggregated network;
    \item a {\sl row-normalised} contact matrix $C$, in which the element $C_{\rm AB}$ gives the probability for a node in group $A$ to have a link to a node of group $B$;
    \item the lists $\{w\}^{\rm int}$ and $\{w\}^{\rm ext}$ of link weights for respectively internal and external links (internal links are links between nodes that belong to the same group, external links are links between nodes from different groups);
    \end{itemize}

  \item we compute the number of links $e$ that must be added to keep the network density constant when we add the $n$ excluded nodes;

  \item we construct $e$ links according to the following procedure:
    \begin{itemize}
    \item a node $i$ is randomly chosen from the set ${\cal P} \backslash {\cal \tilde{P}}$ of excluded nodes;
    \item knowing the group $A$ that $i$ belongs to, we extract at random a target group $B$ with probability given by $C_{\rm AB}$;
    \item we draw a target node $j$ at random from $B$ (if $B=A$, we check that $j \ne i$);
    \item depending on whether nodes $i$ and $j$ belong to the same group or not, we draw from $\{w\}^{\rm int}$ or $\{w\}^{\rm ext}$ 
the weight $w_{ij}$ of the link $ij$;
    \item as for the ${\bf W}$ method, we extract at random $w_{ij}/\Delta t$ contact events of 
length $\Delta t=20s$ within the activity timeline.
    \end{itemize}
  \end{enumerate}
  
\item[WT] We perform the reconstruction using the network density, the distribution of link weights 
and the temporal structure of the contacts given by the distributions of contact durations, inter-contact durations, number of contacts per link and initial waiting times before the first contact,
 all measured in the resampled network ${\cal \tilde{T}}$. The algorithm goes as follows:
  \begin{enumerate}
  \item we compute from the activity timeline the time $T_{\rm u}$ as the total duration of the periods during which contacts can occur.

  \item we measure in the resampled data:
    \begin{itemize}
    \item the density $\rho$ of links in the time-aggregated network;
    \item the list $\{\tau_{\rm c}\}$ of contact durations;
    \item the list $\{\tau_{\rm ic}\}$ of inter-contact durations;
    \item the list $\{p\}$ of numbers of contacts per link;
    \item the list $\{t_0\}$ of initial waiting times before the first contact for each link;
    \end{itemize}

  \item we compute the number of links $e$ that must be added to keep the network density constant when we add the $n$ excluded nodes;

  \item we construct $e$ links according to the following procedure:
    \begin{enumerate}
    \item a node $i$ is randomly chosen from the set ${\cal P} \backslash {\cal \tilde{P}}$ of excluded nodes;
    \item a node $j$ is randomly chosen from the set ${\cal P} \backslash \{i\}$ of all other nodes;
    \item we draw from $\{p\}$ the number of contact events $p$ taking place over the link $ij$;
    \item from $\{t_0\}$, we draw the initial waiting time $t_0$ before the first contact;
    \item from $\{\tau_{\rm c}\}$, we draw $p$ contact durations $\tau_{\rm c}^k$, $k=1,\cdots,p$;
    \item from $\{\tau_{\rm ic}\}$, we draw $p-1$ inter-contact durations $\tau_{\rm ic}^m$, $m=1,\cdots,p-1$;
    \item while $t_0 + \sum_k \tau_{\rm c}^k + \sum_m \tau_{\rm ic}^m > T_{\rm u}$, we repeat steps (c) to (f);
    \item from $t_0$,the $\tau_{\rm c}^k$ and $\tau_{\rm ic}^m$, we build the contact timeline of the link $ij$;
    \item finally, we insert in the contact timeline the breaks defined by the activity timeline.
    \end{enumerate}
  \end{enumerate}

\end{description}

\subsection*{Reconstruction with fixed transitivity}

In order to constrain the transitivity to its value measured in the resampled data, 
we add to the WST algorithm the following elements:

\begin{enumerate}

\item we measure in the resampled data the transitivity $\sigma_0$ of the time-aggregated network;

\item for the construction of each link of a node $i$:
  \begin{itemize}
  \item we calculate the current transitivity $\sigma$ of the network;
  \item we list the potential targets $j$ in two lists $C_\Delta$ and $C_\wedge$, depending on whether the creation of a link between $i$ and $j$ would close a triangle or not;
  \item
    \begin{itemize}
    \item if $\sigma < \sigma_0$, we draw a target node $j$ at random from $C_\Delta$ such that $i$ and $j$ are not linked;
    \item else if $\sigma > \sigma_0$, we draw a target node $j$ at random from $C_\wedge$ such that $i$ and $j$ are not linked.
    \end{itemize}
  \end{itemize}
\end{enumerate}

We show in Supplementary Figure \ref{fig:infer_transitivity} the outcome of simulations performed on reconstructed data sets using this modified algorithm.

\end{document}